\documentclass[twocolumn, astrosymb, twocolappendix, tighten]{aastex631}

\hypersetup{linkcolor=red,citecolor=blue,filecolor=cyan,urlcolor=magenta}
\usepackage{float}
\usepackage{booktabs}
\usepackage{subfigure}
\usepackage{tabularx}

\newcommand{\tcbr}[1]{\textcolor{black}{#1}}

\received{25-Jul-2025}
\revised{15-Oct-2025}
\accepted{27-Oct-2025}

\submitjournal{ApJ}

\shorttitle{Probing multi-wavelength features of \textit{Chandra} sources}
\shortauthors{Kumaran et al.}
\usepackage{amsmath}
\usepackage{multirow}
\usepackage{natbib}
\usepackage{subfigure}
\usepackage{tabularx}

\begin{document}

% \title{Identification of \textit{Chandra} X-ray point sources using machine learning: probing multi-wavelength features}

\title{Explainable machine learning classification of \textit{Chandra}
X-ray sources: SHAP analysis of multi-wavelength features}

\correspondingauthor{Shivam Kumaran}
\email{kumaranshivam57@gmail.com, kumaran@sac.isro.gov.in}

\author{Shivam Kumaran}
\affiliation{Space Applications Centre, Ahmedabad 380015, Gujarat, India}

\author{Samir Mandal}
\affiliation{Indian Institute of Space Science and Technology, 
Thiruvananthapuram 699547, Kerala, India
}

\author{Sudip Bhattacharyya}
\affiliation{Department of Astronomy and Astrophysics, Tata Institute of Fundamental Research, 1 Homi Bhabha Road, Mumbai 400005, India
}

\begin{abstract}

Extensive astronomical surveys, like those conducted with the {\em Chandra} X-ray Observatory, detect hundreds of thousands of unidentified cosmic sources. Machine learning (ML) methods offer an efficient, probabilistic approach to classify them, which can be useful for making discoveries and conducting deeper studies. In earlier work, we applied the LightGBM (ML model) to classify 277,069 {\em Chandra} point sources into eight categories: active galactic nuclei (AGN), X-ray emitting stars, young stellar objects (YSO), high-mass X-ray binaries, low-mass X-ray binaries, ultraluminous X-ray sources, cataclysmic variables, and pulsars. In this work, we present the classification table of 54,770 robustly classified sources (over $3\sigma$ confidence), including 14,066 sources at $>4\sigma$ significance. To ensure classification reliability and gain a deeper insight, we investigate the multiwavelength feature relationships learned by the LightGBM model, focusing on AGNs, Stars, and YSOs. We employ Explainable Artificial Intelligence (XAI) techniques, specifically, SHapley Additive exPlanations (SHAP), to quantify the contribution of individual features and their interactions to the predicted classification probabilities.

Among other things, we find infrared-optical and X-ray decision boundaries for separating AGN/Stars, and infrared-X-ray boundaries for YSOs. These results are crucial for estimating object classes even with limited multiwavelength data. This study represents one of the earliest applications of XAI to large-scale astronomical datasets, demonstrating ML models' potential for uncovering physically meaningful patterns in data in addition to classification. Finally, our publicly available, extensive, and interactive catalogue will be helpful to explore the contributions of features and their combinations in greater detail in the future.

\end{abstract}


\section{Introduction} \label{sec:intro}

The field of astronomy in the modern era has become extremely data-intensive. The large volume of data coming from high-end instruments and serendipitous surveys has made the conventional method prohibitively slow. The use of machine learning (ML) and deep learning (DL) methods is indispensable for analysing and studying these large datasets. Several works over the past decade have established the efficiency, accuracy and competency of ML models for various tasks, including identification and classification of sources \citep{kim2016star}, prediction of parameters for astrophysical objects/models \citep{2024A&A...685A.107M, BH2023MNRAS.520.4867Q}, serendipitous identification of transient events \citep{killestein2021transient}. In the high energy domain, observatories like {\em Chandra X-ray Observatory}, {\em Rossi X-Ray Timing Explorer (RXTE)}, {\em Swift-XRT} and {\em XMM-Newton} have generated a point source catalogue of hundreds of thousands of X-ray objects. 
The latest release from {\it Chandra} is the Chandra Source Catalogue 2.1 \citep{2023HEAD...2040401M}, which contains almost 4,00,000 point sources. These sources consist mainly of Active Galactic Nuclei (AGNs), X-ray emitting stars (hereafter referred to as Stars), Young Stellar Objects (YSOs), X-ray binaries (XRBs), among others. Identifying and classifying the sources becomes a crucial step for various tasks such as target selection, filtering of sources in a selected field, and conducting a statistical population study. Their rigorous classification is done using manual methods such as creating boundaries in color-color diagram \citep{color-color2004ApJ...617..746D}, spectroscopic analysis \citep{kauffmann2003MNRAS.346.1055K}, timing analysis \citep{lin2013classification}. The data table generated through the automated pipeline of the all-sky surveys contains sources' observed properties and simple model-derived parameters. In recent years decision tree based ML models, such as Random Forest \citep{RF2001MachL..45....5B}, Light Gradient Boosted Machine: LightGBM \citep{ke2017lightgbm} have been successfully applied to identify X-ray sources from {\em Chandra, XMM-Newton, SWIFT-XRT}, using sources' properties data-table with significant confidence (LightGBM: \citealt{kumaran2023MNRAS.520.5065K}, Random Forest: \citealt{ml-ex-yang2022ApJ...941..104Y, ml-ex32015ApJ...813...28F}, Unsupervised learning: \citealt{ml-ex5unsupervised-csc}, LogitBoost: \citealt{ml-ex42021MNRAS.503.5263Z}). 

In \cite{kumaran2023MNRAS.520.5065K} (hereafter referred to as \textit{paper-I}), we classified the sources in the Chandra Source Catalogue CSC-2.0 \citep{CSC2024ApJS..274...22E}. We used LightGBM \citep{ke2017lightgbm} as the classifier, and the CSC-2.0 flux, variability properties, along with multi-wavelength data from various observatories. We classified 277069 {\em Chandra} point X-ray sources, of which 54770 (14066) were classified with $3\sigma$ ($4\sigma$) confidence. Although the classification using ML is validated with various methods, the key challenge in the acceptance of ML results for scientific analysis is the black-box nature of these models \citep{BB8237633}. Unlike conventional methods based on physical principles, most ML models, due to their complex and nonlinear architecture, lack direct mechanisms for interpreting learned patterns. Only simpler techniques, like Naive Bayes classifiers and principal component analysis, offer insights that can be translated into human-understandable terms. \cite{XAIfleisher2022understanding} has discussed the importance of transparency, interpretability, and explainability for making the results obtained from ML models trustworthy. To this end, numerous methods have been proposed, collectively referred to as Explainable AI (XAI) \citep{xaidef10.1016/j.inffus.2019.12.012}. \cite{grad-cam8237336} proposed \textit{Grad-CAM}, which attempts to make the deep convolution networks transparent by the visualisation of inner layers via gradient flow. For interpretability, LIME \citep{lime10.1145/2939672.2939778} uses simpler surrogate models to assist local interpretation of the predictions. \cite{shap10.5555/3295222.3295230} introduced a game theory-based method called Shapley Additive exPlanation (SHAP), which borrows the concept of Shapley values from game theory to obtain a local explanation of individual predictions by ML models. A few recent works have demonstrated the use of XAI in astronomy to arrive at an understanding of physical processes. \cite{xaiflare2023A&A...671A..73P} used Grad-CAM to distinguish the Mg II spectra of flaring and non-flaring regions for a model trained to predict solar flares. \cite{BH2023MNRAS.520.4867Q} used SHAP analysis to explain the prediction of black hole parameters from a Random Forest model. \cite{gaiashap2025A&A...697A.107Y} used SHAP values for highlighting the part of stellar spectra responsible for carbon star identification.
  
We use SHAP analysis to provide local explanations for the class membership probabilities of all sources from our previous work. The majority classes: AGNs, YSOs, and Stars show significantly higher global confidence levels, so we focus on them to extract global explanations and feature-importance patterns. We present the classification data for confidently identified sources from {\em paper-I}, along with local explanations for individual predictions, and demonstrate how these local explanations inform classification criteria for AGN, Stars, and YSO sources.

In \S2, we briefly outline the classification methodology used in previous work (referred to as {em paper-1}) to categorize CSC-2.0 sources into eight classes. We also present the classification data-table confidently classified sources, along with their class and class membership probabilities (CMP), with more insight into the CMPs. \S3 introduces the SHAP analysis, and the detailed methodology adopted in this work to use SHAP values for deriving local and global feature importances. In \S4, we present the global explanations and the relation between features' values and their SHAP values. The summary and conclusions are presented in \S5.
\begin{figure*}[]
        \centering   
         \includegraphics[width=1.0\textwidth]{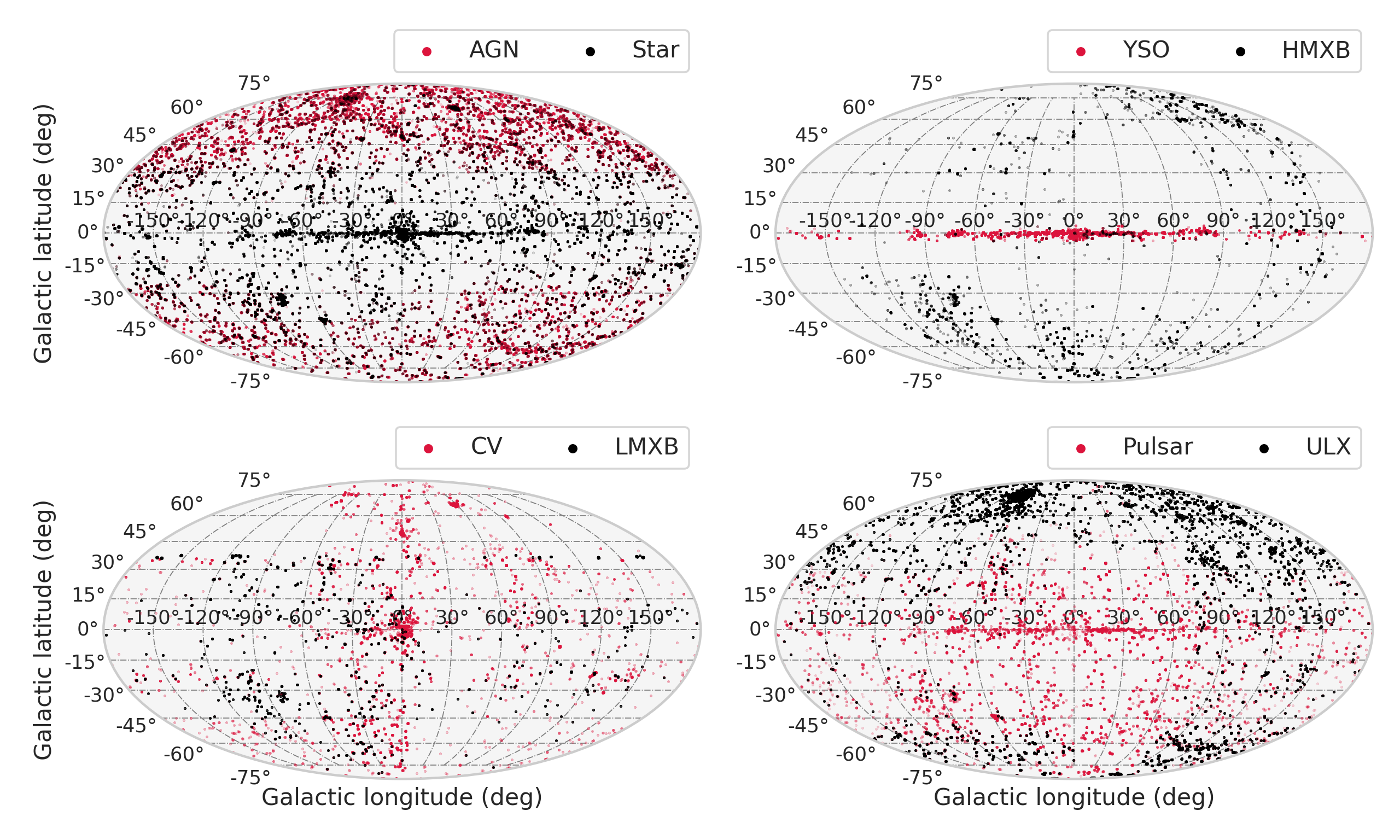}
        \caption{Location of the sources in Galactic coordinates belonging to different classes: AGN, Stars (top-left); YSO, HMXB (top-right); CV, LMXB (bottom-left); Pulsars, ULX (bottom-right) in Aitoff projection. \textit{Note: the pairs for the plot are selected for better visualisation.}}
        \label{fig:source-loc}
\end{figure*}

\section{CSC-2.0 source classification} 
\subsection{Classification Methodology}\label{sec:classification}
In {\em paper-I}, we used the LightGBM model to classify the sources in the Chandra Source Catalogue 2.0 (CSC-2.0). For all the sources, we used the flux values in \textit{Chandra}'s soft (u-csc), medium (m-csc), hard band (h-csc) and broadband (b-csc) along with the inter-observation and intra-observation variability properties as the classification features. \tcbr{To align with other multi-wavelength observations, instead of X-ray fluxes, a proxy of X-ray magnitudes is used by taking the log of observed flux in {\em chandra} bands.} In addition, we also compiled the multi-wavelength features using a conservative cross-match radius of 1 arcsec from {\em Gaia}, {\em 2MASS}, {\em MIPS-Spitzer}, {\em GALEX}, {\em WISE} and {\em SDSS}. We use a total of 41 features (refer to Table 2 of {\em paper-I}) to train the classifier model.
   
We classify the sources belonging to the classes: AGN, Star, YSO, High mass X-ray binaries (HMXB), Low mass X-ray binary (LMXB), ultra-luminous X-ray source (ULX), Cataclysmic variable (CV) and Pulsar. For supervised learning, we prepared a list of confidently identified 7703 objects (AGN: 2395, Stars: 2790, YSOs: 1149, HMXBs: 748, LMXBs: 143; ULXS: 211, CVs: 166 and Pulsars: 101) from various published literature (refer to Table 3 of {\em paper-I}). After doing a comparative study of various decision tree based models, oversampling techniques and imputation methods, we identified that LightGBM with class-weightage and no-imputation gave the best scores. We achieved 93\% precision, 93\% recall, and 0.91 Mathew’s Correlation coefficient (MCC) score.

\subsection{Classification result} \label{sec:result}

We present the distribution of the identified sources in the sky coordinates in Figure \ref{fig:source-loc}. As expected, almost all the AGNs are away from the Galactic plane, and YSOs are on the Galactic plane. Due to reddening by the Galactic plane, the X-ray emission from AGNs drops out of the \textit{Chandra's} regime (However deep {\it Chandra} X-ray surveys have identified significant number of AGNs over typical Galactic plane fields \citep{chandraDeep2005ApJ...635..214E, chandraDeep2009ApJ...701..811T, agnIRexcess2011A&A...526A..86G}. This explains the bias of identified AGNs away from the Galactic plane (top-left panel of Figure \ref{fig:source-loc}). The stars are distributed throughout the sky, most concentrated on the Galactic plane. The HMXBs and ULXs are mostly away from the Galactic plane, indicating that they belong to external galaxies. CVs and Pulsars are mainly concentrated near the Galactic centre.

\begin{table*}
\centering
\caption{A sample of source classification with identified class and associated probabilities. Columns: \textbf{NAME} (Observation ID of the source in the CSC-2.0); \textbf{RA (J2000)}; \textbf{Dec (J2000)}; \textbf{class 1}: Predicted class with highest CMP; \textbf{CMP1}: probability for highest probable class; \textbf{class 2}: Predicted class with second highest CMP; \textbf{CMP2}: probability of second highest class. The complete classification table is available in a machine-readable format at \dataset[DOI:10.5281/zenodo.17346885]{https://doi.org/10.5281/zenodo.17346885} and on the github repository\footnote{\url{https://github.com/KumaranShivam5/Chandra-XAI.git}}} 
\centering
\begin{tabularx}{0.95\textwidth}{cLllccccR} 
\toprule 
\multicolumn{1}{c}{\textbf{Sl No.}} & \multicolumn{1}{c}{\textbf{NAME}} & \multicolumn{1}{c}{\textbf{RA}} & \multicolumn{1}{c}{\textbf{DEC}} & \multicolumn{1}{c}{\textbf{class 1}} & \multicolumn{1}{c}{\textbf{CMP1}} &  \multicolumn{1}{c}{\textbf{class 2}} & \multicolumn{1}{c}{\textbf{\textbf{CMP2}}} \\ \midrule

1 & 2CXO J035844.6+102451 & $03h\, 58m\, 44.69s$ & $+10^\circ\, 4^\prime\, 51^{\prime\prime}.76$ & AGN & 0.964 & LMXB & 0.015 \\
2 & 2CXO J024439.5-593032 & $02h\,44m\,39.52s$ & $-59^\circ\,30^\prime\,32^{\prime\prime}.07$ & AGN & 0.600 & CV & 0.305 \\
3 & 2CXO J014220.8-005331 & $01h\,42m\, 20.81s$ & $-00^\circ\, 53^\prime\,  31^{\prime\prime}.26$ & AGN & 0.997 & STAR & 0.002 \\
4 & 2CXO J042946.9-025027 & $04h\,29m\, 46.98s$ & $-02^\circ\, 50^\prime\,  27^{\prime\prime}.63$ & AGN & 0.619 & STAR & 0.310 \\
5 & 2CXO J150519.5+613017 & $15h\,05m\, 19.56s$ & $+61^\circ\, 30^\prime\,  17^{\prime\prime}.50$ & AGN & 0.987 & ULX & 0.007 \\
6 & 2CXO J052241.4+332050 & $05h\,22m\, 41.49s$ & $+33^\circ\, 20^\prime\,  50^{\prime\prime}.04$ & STAR & 0.995 & YSO & 0.005 \\
7 & 2CXO J231249.0-213414 & $23h\,12m\, 49.02s$ & $-21^\circ\, 34^\prime\,  14^{\prime\prime}.06$ & STAR & 0.989 & AGN & 0.010 \\
8 & 2CXO J171437.2-292735 & $17h\,14m\, 37.22s$ & $-29^\circ\, 27^\prime\,  35^{\prime\prime}.75$ & STAR & 0.919 & PULSAR & 0.041 \\
9 & 2CXO J064149.9-495825 & $06h\,41m\, 49.95s$ & $-49^\circ\, 58^\prime\,  25^{\prime\prime}.45$ & STAR & 0.999 & CV & 0.000 \\
10 & 2CXO J183119.8-020816 & $18h\,31m\, 19.88s$ & $-02^\circ\, 08^\prime\,  16^{\prime\prime}.29$ & STAR & 0.821 & YSO & 0.176 \\
11 & 2CXO J023649.3+593921 & $02h\,36m\, 49.35s$ & $+59^\circ\, 39^\prime\,  21^{\prime\prime}.46$ & YSO & 0.435 & STAR & 0.425 \\
12 & 2CXO J111357.8-611443 & $11h\,13m\, 57.90s$ & $-61^\circ\, 14^\prime\,  43^{\prime\prime}.26$ & YSO & 0.988 & STAR & 0.012 \\
13 & 2CXO J174712.7-282657 & $17h\,47m\, 12.75s$ & $-28^\circ\, 26^\prime\,  57^{\prime\prime}.96$ & YSO & 0.830 & STAR & 0.092 \\
14 & 2CXO J155424.7-551150 & $15h\,54m\, 24.75s$ & $-55^\circ\, 11^\prime\,  50^{\prime\prime}.24$ & YSO & 0.732 & PULSAR & 0.144 \\
15 & 2CXO J131233.5-624216 & $13h\,12m\, 33.56s$ & $-62^\circ\, 42^\prime\,  16^{\prime\prime}.97$ & YSO & 0.919 & STAR & 0.081 \\
16 & 2CXO J010352.8-220815 & $01h\,03m\, 52.80s$ & $-22^\circ\, 08^\prime\,  15^{\prime\prime}.43$ & HMXB & 0.789 & AGN & 0.165 \\
17 & 2CXO J134038.3-313805 & $13h\,40m\, 38.35s$ & $-31^\circ\, 38^\prime\,  05^{\prime\prime}.65$ & HMXB & 0.952 & CV & 0.023 \\
18 & 2CXO J231413.8-423821 & $23h\,14m\, 13.84s$ & $-42^\circ\, 38^\prime\,  21^{\prime\prime}.83$ & HMXB & 0.577 & CV & 0.324 \\
19 & 2CXO J011949.3-411114 & $01h\,19m\, 49.35s$ & $-41^\circ\, 11^\prime\,  14^{\prime\prime}.50$ & HMXB & 0.320 & AGN & 0.313 \\
20 & 2CXO J015116.2-595631 & $01h\,51m\, 16.27s$ & $-59^\circ\, 56^\prime\,  31^{\prime\prime}.11$ & HMXB & 0.244 & LMXB & 0.188 \\
21 & 2CXO J083108.5+523838 & $08h\,31m\, 08.55s$ & $+52^\circ\, 38^\prime\,  38^{\prime\prime}.89$ & LMXB & 0.600 & AGN & 0.377 \\
22 & 2CXO J060232.9+421754 & $06h\,02m\, 32.95s$ & $+42^\circ\, 17^\prime\,  54^{\prime\prime}.91$ & LMXB & 0.551 & AGN & 0.230 \\
23 & 2CXO J203508.1-593628 & $20h\,35m\, 08.12s$ & $-59^\circ\, 36^\prime\,  28^{\prime\prime}.99$ & LMXB & 0.527 & STAR & 0.284 \\
24 & 2CXO J002346.1-720024 & $00h\,23m\, 46.14s$ & $-72^\circ\, 00^\prime\,  24^{\prime\prime}.94$ & LMXB & 0.448 & CV & 0.284 \\
25 & 2CXO J015744.9+374439 & $01h\,57m\, 44.93s$ & $+37^\circ\, 44^\prime\,  39^{\prime\prime}.59$ & LMXB & 0.515 & CV & 0.163 \\
26 & 2CXO J114617.8+202248 & $11h\,46m\, 17.83s$ & $+20^\circ\, 22^\prime\,  48^{\prime\prime}.95$ & ULX & 0.641 & AGN & 0.180 \\
27 & 2CXO J065105.1+412949 & $06h\,51m\, 05.15s$ & $+41^\circ\, 29^\prime\,  49^{\prime\prime}.29$ & ULX & 0.568 & PULSAR & 0.238 \\
28 & 2CXO J122501.5+125236 & $12h\,25m\, 01.57s$ & $+12^\circ\, 52^\prime\,  36^{\prime\prime}.14$ & ULX & 0.846 & AGN & 0.120 \\
29 & 2CXO J150640.0+013352 & $15h\,06m\, 40.03s$ & $+01^\circ\, 33^\prime\,  52^{\prime\prime}.02$ & ULX & 0.340 & PULSAR & 0.322 \\
30 & 2CXO J192008.9+440359 & $19h\,20m\, 09.00s$ & $+44^\circ\, 03^\prime\,  59^{\prime\prime}.63$ & ULX & 0.636 & CV & 0.167 \\
31 & 2CXO J115112.2-284649 & $11h\,51m\, 12.25s$ & $-28^\circ\, 46^\prime\,  49^{\prime\prime}.29$ & CV & 0.573 & PULSAR & 0.187 \\
32 & 2CXO J180434.1-281850 & $18h\,04m\, 34.19s$ & $-28^\circ\, 18^\prime\,  50^{\prime\prime}.26$ & CV & 0.988 & HMXB & 0.007 \\
33 & 2CXO J125303.8-292758 & $12h\,53m\, 03.87s$ & $-29^\circ\, 27^\prime\,  58^{\prime\prime}.74$ & CV & 0.275 & PULSAR & 0.238 \\
34 & 2CXO J234131.1-540855 & $23h\,41m\, 31.12s$ & $-54^\circ\, 08^\prime\,  55^{\prime\prime}.72$ & CV & 0.385 & LMXB & 0.198 \\
35 & 2CXO J132510.8-425214 & $13h\,25m\, 10.87s$ & $-42^\circ\, 52^\prime\,  14^{\prime\prime}.53$ & CV & 0.820 & STAR & 0.145 \\
36 & 2CXO J201828.8+113527 & $20h\,18m\, 28.81s$ & $+11^\circ\, 35^\prime\,  27^{\prime\prime}.30$ & PULSAR & 0.422 & AGN & 0.322 \\
37 & 2CXO J124853.7-411816 & $12h\,48m\, 53.80s$ & $-41^\circ\, 18^\prime\,  16^{\prime\prime}.96$ & PULSAR & 0.522 & STAR & 0.422 \\
38 & 2CXO J231103.5-214714 & $23h\,11m\, 03.58s$ & $-21^\circ\, 47^\prime\,  14^{\prime\prime}.09$ & PULSAR & 0.977 & AGN & 0.011 \\
39 & 2CXO J204341.6+170842 & $20h\,43m\, 41.65s$ & $+17^\circ\, 08^\prime\,  42^{\prime\prime}.10$ & PULSAR & 0.950 & STAR & 0.022 \\
40 & 2CXO J015454.7-554200 & $01h\,54m\, 54.70s$ & $-55^\circ\, 42^\prime\,  00^{\prime\prime}.96$ & PULSAR & 0.543 & STAR & 0.156 \\
\bottomrule
\end{tabularx}%
    \label{tab:source-table}
\end{table*}
With the LightGBM classifier, we assign class membership probabilities (CMP) for each object corresponding to each of the eight classes. The class assigned to the source is the one with the highest CMP. \tcbr{In {\em Paper I} we adopted the terminology of $>3\sigma$ and $>4\sigma$ to denote CMP thresholds, by analogy to conventional confidence levels. However, unlike Gaussian statistics, these thresholds should not be interpreted as formal significance levels. They represent probability cutoffs derived from the machine learning classifier, whose reliability is best assessed using performance metrics such as precision, recall, f1-score and MCC. Throughout this work, we therefore treat the thresholds as relative measures of classification robustness, rather than exact statistical confidences}. Table \ref{tab:source-table} gives a list of some selected samples for each class. The source name (ID), position on the sky, the highest and second highest probable class and their respective CMPs are indicated. The purpose of this table is not to disclose the names of the sources identified with the highest confidence, but rather to present a randomly selected subset for the purpose of discussing certain issues related to classification. It also highlights the relevance of referencing CMP1 and CMP2 in this context.

\begin{figure}[ht!]
        \centering
        \includegraphics[width=\columnwidth]{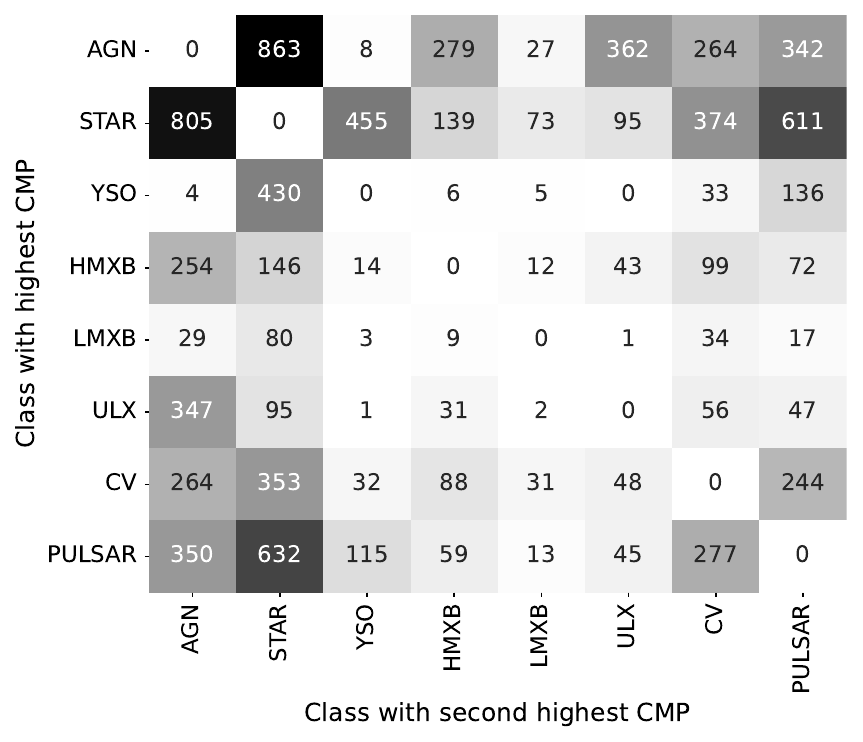}
        \caption{Confusion matrix corresponding to the confused sources (CMP1-CMP2 $\leq 0.05$) in the identified data set. The Y-axis shows the highest probable class, i.e., \textbf{class 1} column in Table~\ref{tab:source-table}, and the X-axis is the second highest probable class (\textbf{class 2} column in Table~\ref{tab:source-table}).}
        \label{fig:conf-src}
    \end{figure}

In some cases, the CMP1 value may not be particularly high, which might suggest a lack of confidence in the classification. However, a significantly lower CMP2 value reinforces the reliability of the classification by providing a strong contrast between the top candidates. For example, source \textit{2CXOJ015744.9+374439} (SI No. 25 in Table \ref{tab:source-table}) has a CMP1 value of 0.515 for an LMXB class and a CMP2 value of 0.163 for a CV, making the classification clearly favour an LMXB. In contrast, when the highest and second-highest CMP values are close, for example, source \textit{2CXOJ124853.7-411816} (SI No. 37 in Table \ref{tab:source-table}) with CMP1 = 0.522 for Pulsar and CMP2 = 0.440 for Star, the classification remains uncertain, as the source shows comparable likelihood for both classes. Out of the 269366 newly classified sources, only 9254 sources have the difference between the probabilities of the top two classes CMP1-CMP2 $<0.05$. Figure \ref{fig:conf-src} shows the confusion matrix of all such sources. The highest confusion is mostly between AGNs and stars. Due to majority bias, most of these cases are confused with the majority class.  ULXs and HMXBs are mainly confused with AGNs. Pulsars are mostly confused with stars. However, significant cases of pulsars are equally confused with CVs and YSOs. \tcbr{Although the primary focus of this study is on confidently classified sources, the Figure \ref{fig:conf-src} illustrates the confusion patterns among sources with  CMP1-CMP2 $<0.05$. These ambiguous cases highlight where the model encounters difficulty in separating classes (e.g., AGN vs. Star), and they motivate the subsequent SHAP-derived thresholds that improve interpretability of the decision boundaries. Thus, \ref{fig:conf-src} provides useful context for understanding why certain features become critical for classification.} 

    \begin{figure}[ht!]
        \centering
        \includegraphics[width=\columnwidth]{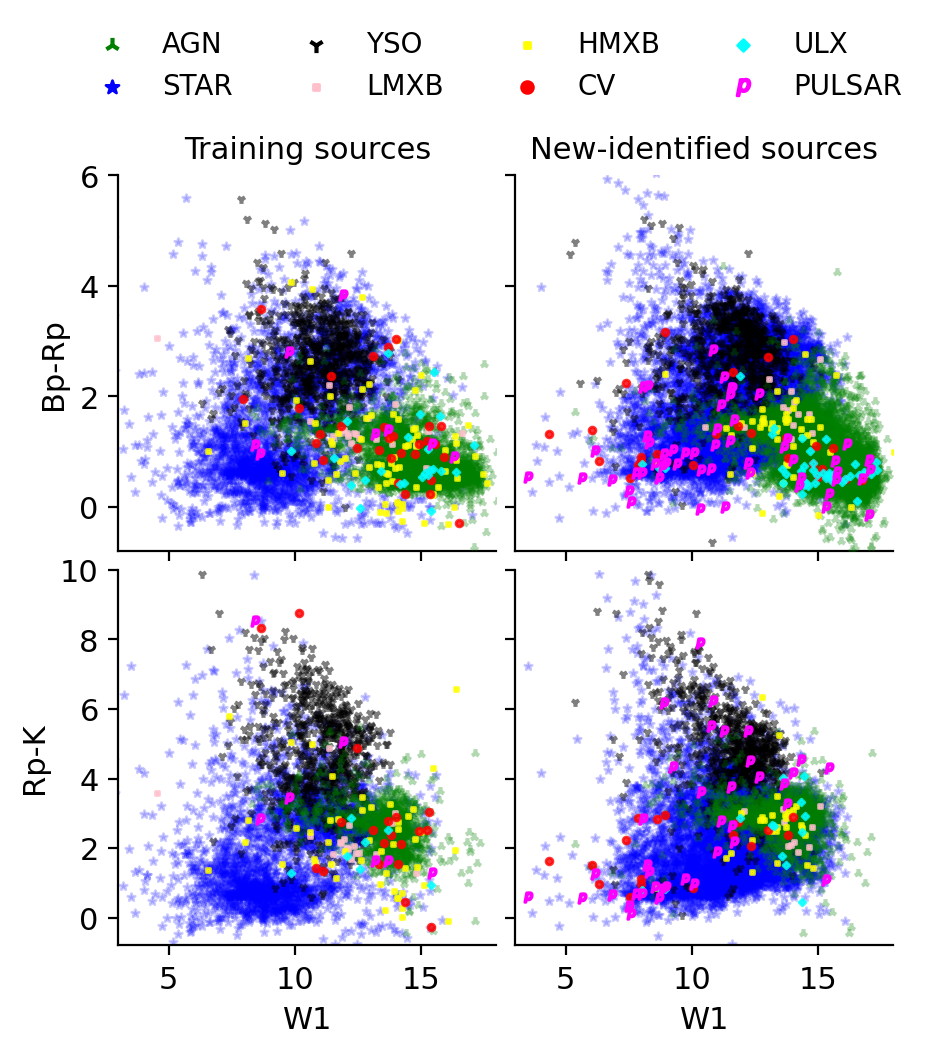}
        \caption{Comparison of the distribution of sources in the training dataset to the sources in the newly identified dataset on the Optical ({\em Gaia}) and IR ({\em WISE} and {\em 2MASS}) color magnitude diagram.}
        \label{fig:TD-cluster-comparison}
    \end{figure}
    For a visual comparison of newly classified sources, we investigate the source properties distribution in arbitrarily selected feature space shown as scatterplot in Figure \ref{fig:TD-cluster-comparison}. We observe that the classified sources generally follow the trends seen in the feature–feature space of the training sample, lending support to the overall reliability of the classification results. While the choice of features shown in these illustrative plots is arbitrary, and thus not suitable for drawing quantitative conclusions about feature importance, it is important to note that manual analysis is impractical due to the high dimensionality of the data, since 41 features and their mutual interactions impact the classification result. This limitation motivates the investigation of which features play the most significant roles in classification, identify any meaningful thresholds, and explore potential clustering of different source classes in feature space using advanced machine learning techniques.

\section{SHAP analysis for classification explanation} \label{sec:shap}
    \subsection{Principle: Shapely values}
     
     \begin{figure*}
        \centering
        \includegraphics[width=\textwidth]{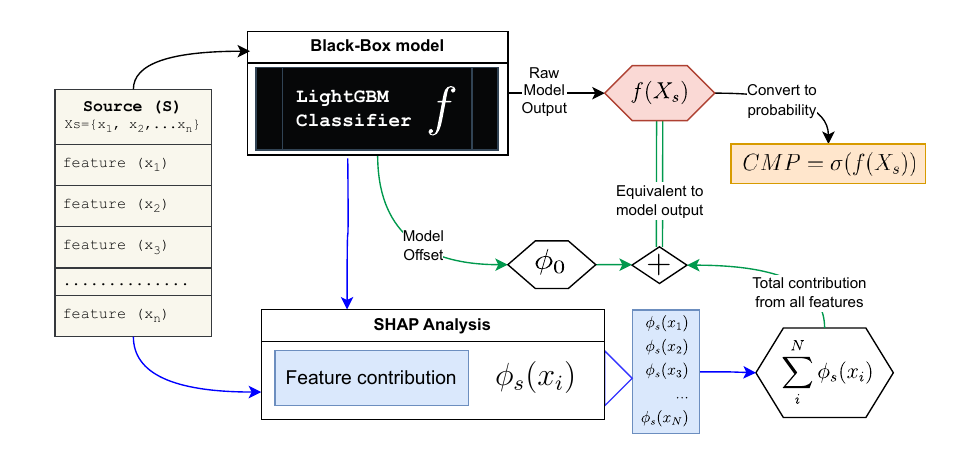}
        \caption{General workflow for local explanation of a black-box classifier model output using SHAP Analysis. Starting with the source's feature set $X_s$, the LightGBM model $f$ generates a raw model output $f(X_s)$ which is converted to CMP for the given class using the sigmoid ($\sigma$) function. This chain is indicated by black arrows. The blue arrows show the workflow of allocating the model output, $f(X_s)$, to the feature contributions using SHAP analysis. SHAP Value for each feature, $\phi_s(x_i)$, is calculated from Equation~\ref{eq:shap-calc}. The green arrows outline equivalence between features' SHAP values and the model raw output $f(X_s)$ with additional model offset $\phi_0$. See \S\ref{sec:shap} for details.}
        \label{fig:shap-analysis}       
    \end{figure*}   
    
   Our goal is to understand why our LightGBM model predicts a certain class membership probability (CMP) for a specific source. We want to see how each individual feature contributes to that particular prediction. This is referred to as a local explanation because it focuses on a single instance. From the statistics for many local explanations, we can then understand the overall importance of features. 

   To achieve this, we use the \textbf{Shapely Additive exPlanations (SHAP)} \citep{lundberg2017unified} analysis technique. SHAP borrows ideas from cooperative game theory, treating each input feature as a `player' in a game. The `reward' in this game is the model's raw output for a given object, and SHAP fairly distributes this reward among the features based on their individual contributions. 

   In mathematical notation, the LightGBM model can be defined as a function $f$ that maps a set of MW feature values $X_s$ to a real number: $f:\{X_S \}\rightarrow \mathcal{R}$. For each object, our LightGBM model (hereafter referred to as $f$) takes 39 multiwavelength (MW) properties as input. The model raw output, $f(X_s)$ is then converted into the CMP using a sigmoid function: $CMP = \sigma(f(X_s))=[1+e^{-f(X_s)}]^{-1}$ for a specific class, i.e., AGN, Star and YSO.

   A feature's SHAP value, $\phi_s(x_i)$, represents its precise contribution to this raw model output.
   The SHAP value for a feature $x_i$, for a specific source $s$ is given by:

    \begin{align}
        \phi_s(x_i) = \sum_{Z} W(Z)\left[f(Z\cup\{x_i\})-f(Z)\right], 
        \label{eq:shap-calc}
    \end{align} where
    \begin{align*}
        W(Z) = \frac{|Z|!(|X_s|-|Z|-1)!}{|X_S|}.
    \end{align*}
    Here, $X_s$ is the set of all features for the source $s$, $Z$ represents all possible subsets of $X_s$, for example $\{x_1\}, \{x_1,x_2\}$, and so on, with the condition that none of these subsets include the feature $x_i$. The subset containing only the feature $x_i$ is represented with $\{x_i\}$. The first term in square brackets of Equation~\ref{eq:shap-calc}, $f(Z\cup\{x_i\})$, represents the model output when the ${x_i}$ feature is included, and the second term, $f(Z)$, is the model output without ${x_i}$. The weight factor $W(Z)$ is the probability of the feature $x_i$ to join a coalition of all possible subsets of $Z$.

    For a balanced dataset and ideal classifier, the expectation value of the model output over all the sources should be unity, i.e., $E[f(X_s)]\forall s$=1, and hence CMP$=0.5$. In this case, the raw model output should be comprised only of the contribution from all the features such that $f(X_s) = \sum_i^N \phi_s(x_i)$. Given the model is trained on real data, it has a non-zero offset, denoted as $\phi_0$ and the relation between the raw model output and the features' SHAP values is:
    \begin{align}
        f(X_s) = \phi_0 + \sum_i^N \phi_s(x_i),
    \end{align}
    where $\phi_0$ is the average raw output of the model across the entire dataset. In simple terms, SHAP values tell us how much each feature pushes the prediction away from the average prediction, allowing us to pinpoint which features are most responsible for a specific classification. \\ \\

    \begin{figure}
        \centering
        \includegraphics[width=0.5\textwidth]{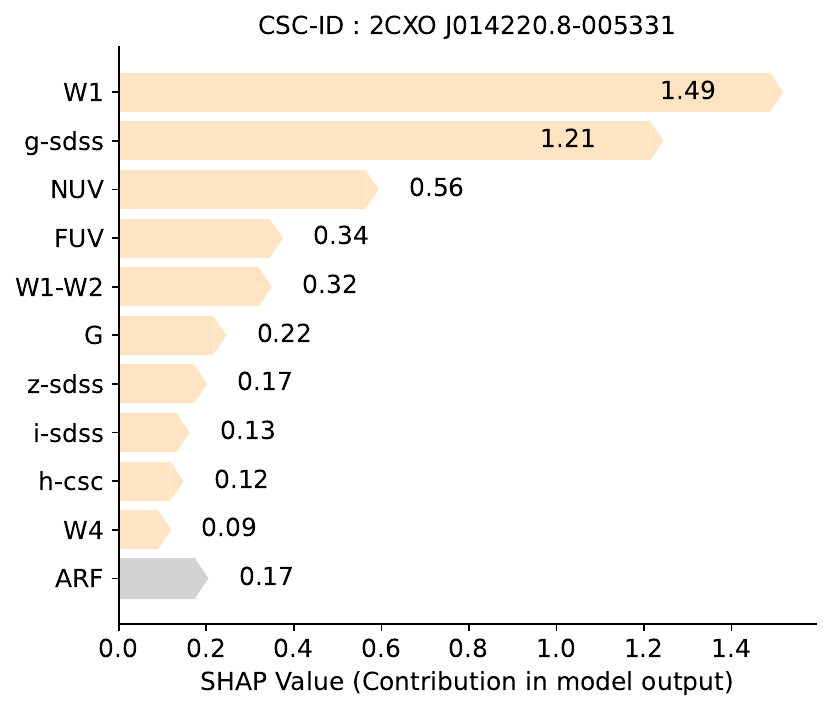}
        \caption{Local explanation for the source \textit{2CXO J01422.8-005331}. The raw output of the LightGBM model for this source is $f(X_s)=4.83$. The individual features SHAP values $\phi_s({x_i})$ are given on the X-axis for the top 10 features $x_i$ on the Y-axis. The sum of contributions from all remaining features is indicated with the `ARF' label in the last row.}
        \label{fig:force-plot}
    \end{figure}
    
    \subsection{Methodology}\label{sec:method}
    The number of sources in the HMXB, LMXB, CV and Pulsar classes is small for a significant statistical study. \tcbr{While {\em paper-I} employed a single multi-class LightGBM classifier, in this work we adopt a one-vs-rest binary formulation for SHAP analysis of the majority classes (AGN, Star, YSO). The one-vs-rest approach isolates the feature contributions that specifically distinguish a given class from all others, thereby providing class-wise interpretability. This enables clearer identification of physically meaningful thresholds and feature interactions. Importantly, the decision surfaces and classification probabilities remain consistent with those obtained in the multi-class methodology\citep{allwein2000reducing, rifkin2004defense}, ensuring that the explanations derived here are directly relevant to the classifications in {\em paper-I}. Other methodological difference from {\em paper-I} in this work is that we exclude the Galactic coordinates from the feature set, as these strongly bias the YSO classification by encoding spatial clustering rather than intrinsic source properties. Since our focus here is on probing the physical roles of multi-wavelength features, we restrict the analysis to photometric and X-ray features}.  Figure \ref{fig:shap-analysis} shows the flowchart for the overall methodology. For getting class-wise SHAP analysis, we implement three binary LightGBM classifiers, one for each class: AGN, Star and YSO. For each class, say `AGN', we prepare balanced training data by including all samples labelled `AGN' and an equal number of samples from other classes and label them as `Non-AGN'. We then train a LightGBM model on this dataset to perform binary classification, assigning a probability of an object being an AGN. The LightGBM raw outputs are converted to a probability using its inbuilt \textsc{sigmoid} function. To determine the contribution, or `share', of each feature to this raw output $f(X_s)$, we use the \textsc{TreeExplainer} \citep{lundberg2020local2global} sub-routine from the \textsc{SHAP} package \citep{NIPS2017_8a20a862}. For a dataset of N sources and M features, the resulting SHAP table will be an N×M matrix, where each cell represents a SHAP value. The sum of the SHAP values along any row of this table (plus the model offset) equals the model's raw output for that specific source.

    One can visualise the local explanation of prediction for an individual source in Figure \ref{fig:force-plot}. It shows an example output prediction of SHAP analysis for the object \textit{2CXO J01422.8-005331} (SI No. 3 in Table \ref{tab:source-table}), classified as an AGN.

     The Y-axis lists important features, and the X-axis shows their individual contributions (may be positive or negative) to the SHAP value. The yellowish arrows indicate the contribution of the top 10 features that are pushing the output of the model towards a positive value. The grey arrow represents the contributions from all other features (ARF), which is the sum of many small positive and negative contributions. The size of the arrow indicates the relative importance of the feature. For the AGN one-vs-rest classifier, the model offset is $\phi_0 = 0.083$. In this example, the model output is $f(X_{s})=4.92$, which is equal to the sum of the model offset ($\phi_0=0.083$) and the total SHAP value $\sum \phi_s{x_i}=4.83$. It corresponds to an AGN class membership probability $P_{AGN}(s)=\sigma(4.92) = 0.993$. Notice that $P_{AGN}(s)$ is slightly different from CMP1 of \textit{2CXO J01422.8-005331} (SI No. 3 in Table \ref{tab:source-table}) because CMP1 is calculated for a multi-class scenario, whereas $P_{AGN}(s)$ is for binary classification. However, the local explanation of top features remains valid for the sources under the majority classes. Using this method, we calculate the contribution made by each feature to the classification of individual sources. In the presented catalogue\footnote{\url{https://github.com/KumaranShivam5/Chandra-XAI.git}}, contribution of feature alongwith the CMPs are given for each source

\section{Result and Discussion}
    Using the LightGBM classifier, we identified the class of 54,770 sources with more than a CMP $>3\sigma$. Being the majority class, AGNs, YSOs and Stars were identified with relatively higher CMP. In this section, we present the result of SHAP analysis to understand the influence of features on the prediction for these majority classes.
  
 \subsection{Global feature importance}
   Conventional global feature importance methods for decision tree (DT) models rely on how often a feature is used in split at a DT branch across all trees in the ensamble (Gini importance) or by using permute-and-predict (PaP)\citep{RF2001MachL..45....5B}. PaP works by shuffling the values of one feature and measuing how much the model's performance drops compared to the original data. Since these feature importance values are based on overall statistics from the validation set, they provide a global understanding of feature influence but can not explain individual predictions. A major drawback of PaP is the assumption that features are independent which is not true for MW features. \cite{strobl2008conditional} highlights this drawback and suggests an improvement by using a conditional permutation scheme.

    The SHAP analysis \citep{lundberg2017unified} overcomes these limitations by making predictions using all possible combinations of features to the already trained model for individual sources without retraining the model on the modified dataset. For a detailed explanation of the methods, including but not limited to random-permutation, gini index, and \textsc{TreeExplainer}, refer to \cite{lundberg2020local2global}. We extract the local explanations of all the sources for AGNs, Stars and YSOs using SHAP analysis of the one-vs-rest classification strategy as described in \S\ref{sec:method}. 
    
    \begin{figure}
        \centering
        \includegraphics[width=0.5\textwidth]{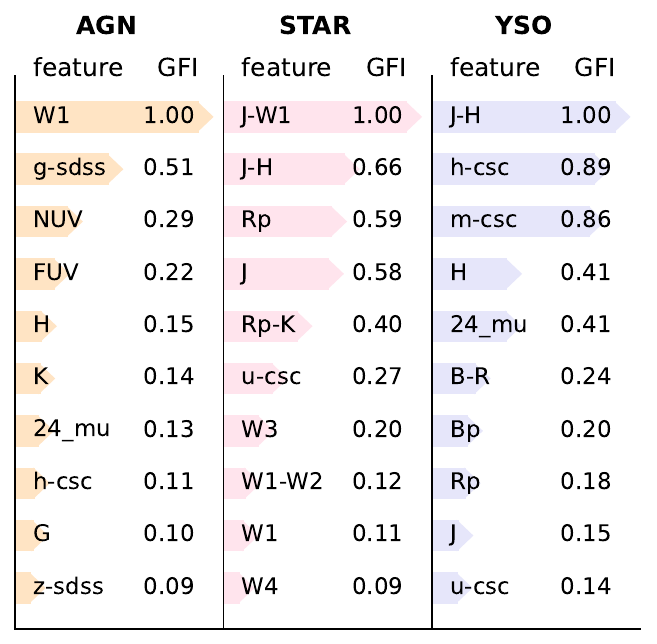}
        \caption{Class-wise Global feature importance (GFI) indicating the relative feature importance for AGN, Stars, and YSOs. The GFI is the expected SHAP values normalized relative to the highest expectation value within each class. See text for details.} 
        \label{fig:feat-imp}
    \end{figure}

    Figure \ref{fig:feat-imp} shows the relative global feature importance (GFI) for the top 10 features of each class.  We are interested in the features that add to the positive prediction of a given class. Therefore, for global SHAP statistics of a given class, we consider only those sources (in a one-vs-rest classifier) which are classified to have CMP$ >0.5$. We then calculate the SHAP distribution of features for all sources in each class.

    The cumulative patterns arising from the local explanations are used to extract the model's global behavior \citep{SHAPexample2025A&A...697A.107Y, BH2023MNRAS.520.4867Q}. \cite{lundberg2020local2global} used the expectation value of SHAP absolute magnitude over all the samples for a given feature as its GFI. Most of the SHAP histograms across all three classes deviate from a normal distribution. Considering this, we measure the expectation value by calculating the probability density function (PDF\footnote{The PDF is calculated using the kernel density estimation \citep{scot1992multivariate} method, implemented in \textsc{Python Scipy} library \url{https://docs.scipy.org/doc/scipy/reference/generated/scipy.stats.gaussian_kde.html}}) of individual feature SHAP value, $\phi_s(x_i)$, across all sources with CMP$ >0.5$.   
    Using these PDFs, we calculate the feature SHAP expectation value, $E[\phi_s(x_i)]$, and normalize by the maximum value. The normalized value, $E[\phi_s(x_i)]/{\rm max}(E[\phi_s(x_i)])$, within the class is presented as the GFI in Figure \ref{fig:feat-imp}. It must be noted that these GFI values are the combined effect of the feature on its own as well as its interaction with other features (refer \S\ref {sec:feature-interaction}). For AGN, the {\em WISE} W1 and {\em SDSS} g band magnitudes play the most important role. This result is consistent with previous identification of AGNs using SDSS g band and various other WISE color criteria \citep{agnWise2015ApJS..221...12S}. The UV features ({\em GALEX}'s FUV and NUV) are important for identifying AGNs from other classes, as they appear only in the AGN top feature list. For Stars apart from the Galactic coordinates (not considered here), the IR (J, H, W1) colors and Optical (Gaia Rp) magnitude are important. The IR excess \citep{ysoIRExcess2013AJ....145..126H} in YSOs results in the highest GFI (J-H colour). The X-ray features (h-csc and m-csc band fluxes) are also important, along with the IR color for YSO.
    
    This GFI list reflects the overall trend in feature contribution calculated as the expectation value of feature importance across all sources. It is important to note that the relative importance of any given individual source may deviate from this global average, as it depends on the source's specific feature value. In the next section, we discuss how the actual value of the feature impacts its importance in the classification.

    \subsection{Relation of feature importance with their values}\label{sec:shap-feat-corr}
    
       \begin{figure*}
        \centering
        \includegraphics[width=\textwidth, ]{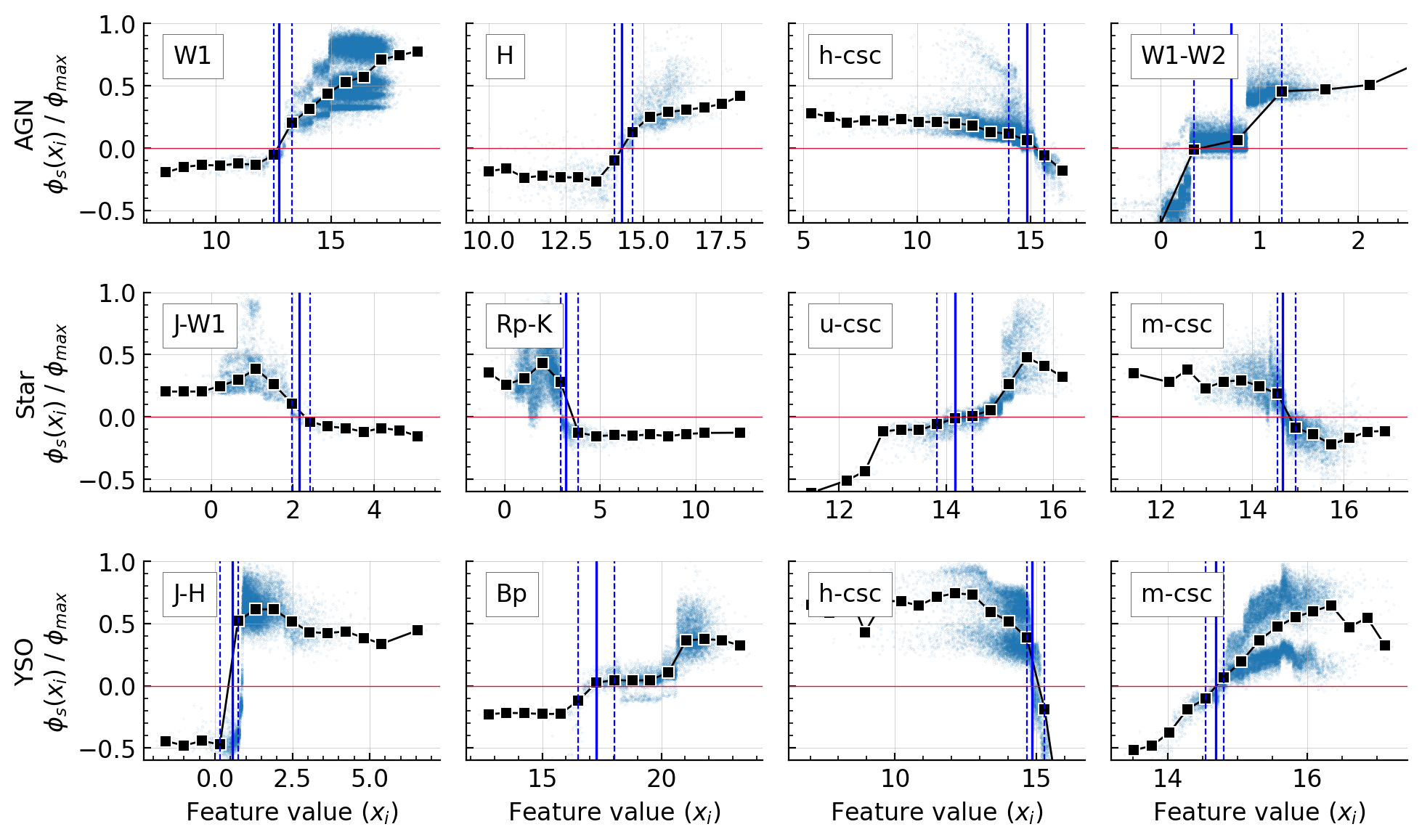}
        \caption{Variation of normalised SHAP values for a few selected features ($x_i$). The blue scatterplot corresponds to individual sources. The scatter points are grouped in 15 equal bins on the X-axis. The black square points are the average value along the Y-axis in each bin and are placed at the center of the bins on the X-axis. The solid blue vertical line shows the threshold value where the trend line (black curve) crosses the zero SHAP value (shown with red line) on the Y-axis. The error in the threshold is shown with vertical blue dashed lines. See text for details.}
        \label{fig:shap-feat-corr}
    \end{figure*}
    
We analyze the variation of local SHAP values to examine how a feature’s importance evolves with its value. This approach allows us to identify features that exhibit a significant correlation between their numerical values and their SHAP contributions to the model's output. In particular, we focus on features that display a threshold value above or below which they contribute systematically and positively (or negatively) to the classification probability. Figure \ref{fig:shap-feat-corr} illustrates four representative examples, each showing a significant positive or negative correlation between feature value and SHAP impact across different classes.

    \begin{table}[]
    \centering
    \caption{The threshold values above or below which the feature has a positive contribution in the respective class identification based on the SHAP-feature correlation from Figure \ref{fig:shap-feat-corr}. The last column gives the percentage of sources in the training data belonging to the respective class meeting the threshold criteria.}

    \begin{tabularx}{\columnwidth}{lccl}
    \toprule
        Class & Feature & Threshold &  Sources (\%) \\  \midrule
        \textbf{AGN} & W1           & $>12.5_{-0.8}^{+0.8}$ & 94.5\\
            & H            & $>14.1_{-0.7}^{+0.5}$ & 86.2\\
            & h-csc        & $<15.1_{-0.2}^{+0.6}$  & 90.0\\
            & W1-W2        & $>0.6_{-0.3}^{+0.2}$ & 73.6\\ 

        \midrule 

        \textbf{Star} & J-W1          & $<2.3_{-0.3}^{+0.1}$ & 85.8\\
            & Rp-K          & $<3.5_{-0.5}^{+0.4}$  & 82.2\\
            &  u-csc         & $>14.1_{-0.2}^{+0.4}$ & 74.5\\
            & m-csc          & $<14.8_{-0.2}^{+0.2}$  & 61.4\\

        \midrule 

        \textbf{YSO} & J-H          & $>0.4_{-0.2}^{+0.3}$ & 98.3\\
            & Bp            & $>17.1_{-0.5}^{+0.2}$ & 89.4\\
            & h-csc         & $<15.0_{-0.4}^{+0.3}$ & 87.7\\
            & m-csc         & $>14.6_{-0.3}^{+0.2}$ & 89.0\\ 
        \bottomrule
    \end{tabularx}%

    \label{tab:feature-threshold-table}
    \end{table}
    Each blue scatter points in the plot represents a source, with the X-axis showing the feature's value ($x_i$) and the Y-axis showing the corresponding normalised (by the maximum value) impact on the output. In each class, we have selected sources having the raw model output $f(X_s) > 1$ such that the CMP $>0.5$. The black squares represent the binning of the scatter points on the X-axis (15 bins), and the corresponding average SHAP value of sources in the given bin. Therefore, black squares show the overall trend and allow for computing the feature threshold value above (or below) which the feature has a positive impact on the model output. We find this threshold (solid blue line) by computing the zero crossing of this line (black line in the figure \ref{fig:shap-feat-corr}) with respect to the Y-axis. The uncertainties (denoted by blue dotted lines) in the threshold values are taken as the bin size at the zero-crossing point in Figure \ref{fig:shap-feat-corr}. However, we quote two bin-widths as the error in the threshold if the crossing point is very close to the bin center, e.g h-csc and W1-W2 for AGN, u-csc for Star and Bp for YSO.

    The top GFI (see Figure~\ref{fig:feat-imp}) in each class contribute more on the positive side beyond a threshold, and it justifies their importance. For AGN, the \textit{Chandra's} h-csc band magnitude $<15.1$ has a positive contribution towards AGN's probability. In other words, higher flux in the hard X-ray band results in a given object being more likely to be classified as AGN. We observe a positive correlation for W1-W2 $>0.6$ for AGNs. The result agrees with the work by \cite{2013ApJ...772...26A}, where they have shown that AGNs can be identified with 90\% reliability using the W1-W2 color-magnitude diagram for candidate AGNs with W2 ($4.6\micron$) magnitude $<11.7$. For stars Rp-K and J-W1, have positive SHAP values, which, after passing the threshold, become negative but close to 0. Although the m-csc feature has an obvious zero crossing point, allowing to get a clear threshold, it also has an equal positive and negative SHAP values distribution, bringing its GFI close to 0. For YSO, h-csc and m-csc show very strong positive contributions beyond the threshold and are placed toward the top of GFIs. The threshold value along with the error estimates for all the cases given in this figure is provided in the Table \ref{tab:feature-threshold-table}. For validation of these thresholds, we verify them against our training dataset. For each class, the `Sources (\%)' column in Table \ref{tab:feature-threshold-table} shows the fraction of total sources in the training dataset (2395 AGNs, 2790 Stars and 1149 YSOs) following this threshold. A very high fraction of  AGN sources follow the W1, H and h-csc thresholds criteria (94.5\%, 86\%, and 90\%  respectively). W1-W2 is followed by relatively fewer, 73.6\%  of sources. For Stars, J-W1 and Rp-K have more than 80\%  sources agreeing with the threshold. The other two features, u-csc and m-csc, have 74\% and 61\% agreement with the training set. For YSO, all the criterion is satisfied by the training data with agreement $> 89\%$. J-H criterion is most confident with 98\% of training sources following this criterion.

     \begin{figure*}
        \centering
        \includegraphics[width=\textwidth]{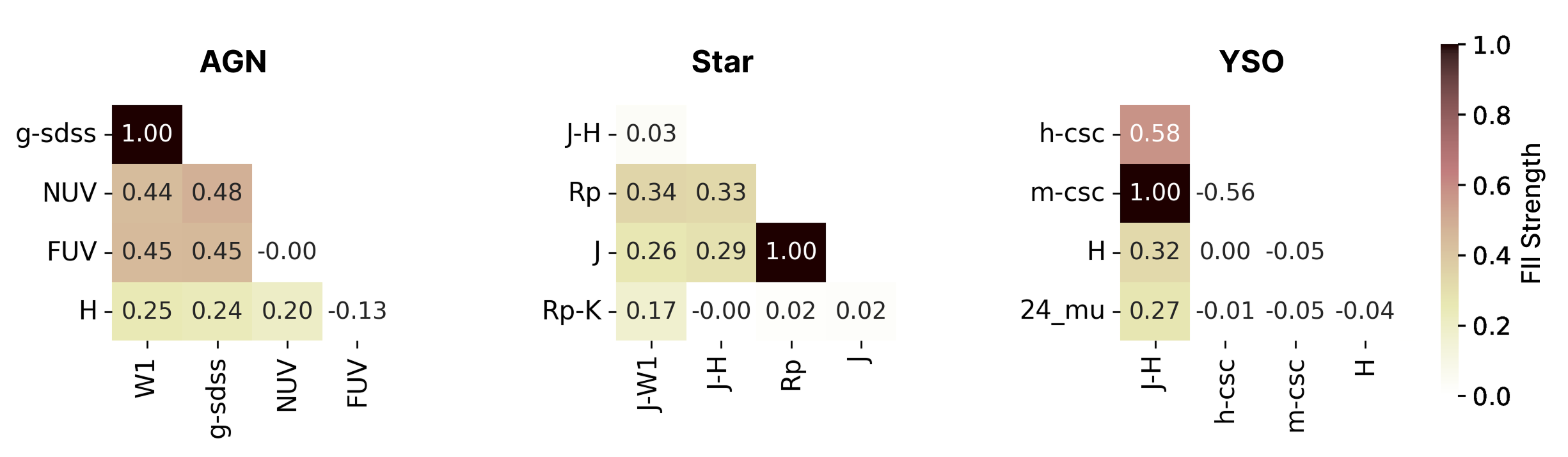}
        \caption{Relative Feature-feature interaction (FFI) between top five GFI (Figure~\ref{fig:feat-imp}) is shown for AGNs (left), Stars (centre) and YSO (right). Each FFI strength is normalised and represented by colour code.}
        \label{fig:feat-int}
    \end{figure*} 
    
    \subsection{Contribution of feature-feature interaction in classification}\label{sec:feature-interaction}

    Inference from SHAP-feature correlation (Figure \ref{fig:shap-feat-corr}) must be carefully derived. The correlation may be a result of a confounding effect. The importance of one feature may be influenced by its interaction with other features as well.

     For the top five features for all the classes, we find their corresponding feature-feature interaction importance (hereafter referred to as FFI) and are denoted as $\phi_s(x_i:x_j)$ for features $x_i$ and $x_j$. The FFI between feature $x_i$ and feature $x_j$ of the source ($s$) is given as: 
    \begin{align}
        \phi_s(x_i:x_j) = \sum_{Z} W(Z)\Delta_{ij}(Z),
        \label{eq:shap-int-calc}
    \end{align}
    where
    \begin{equation*}
        \Delta_{ij}(Z) = [f(Z\cup\{x_i,x_j\})-f(Z\cup\{x_j\})] - [f(Z\cup \{x_i\}) - f(Z)]
    \end{equation*} and
    \begin{align*}
        W(Z) = \frac{|Z|!(|X_s|-|Z|-2)!}{ 2(|X_S|-1)! }
    \end{align*}
    where $X_s$ is the set of all features for the source $s$, $Z$ represents subsets of the features that do not include $x_i$ and $x_j$. The interaction importance between features $x_i$ and $x_j$ is defined as the difference between the SHAP value of $x_i$ when $x_j$ is present (first square bracketed term) and the SHAP value of $x_i$ in the absence of $x_j$ (term in the second square bracket).
    
    For M number of sources with N features, the interaction values are represented in a 3-dimensional data cube of size (M, N, N), where the cell (s, i, j) represents the interaction value between the features $x_i$ and $x_j$ for the $s^{th}$ source. Global feature interaction between features $x_i$ and $x_j$ is calculated by taking the expectation value of the data cube across the first dimension (M). The resultant matrix is the global feature-interaction matrix. A subset matrix for the top five features for each class is given in Figure \ref{fig:feat-int}. FFI strength is normalised by the highest value in the $39\times39$ feature interaction matrix for a given class computed across all the sources with $CMP >0.5$. Note that feature self-interaction does not carry any meaningful insight and hence has not been considered.

     \begin{figure*}
        \centering
        \includegraphics[width=\textwidth]{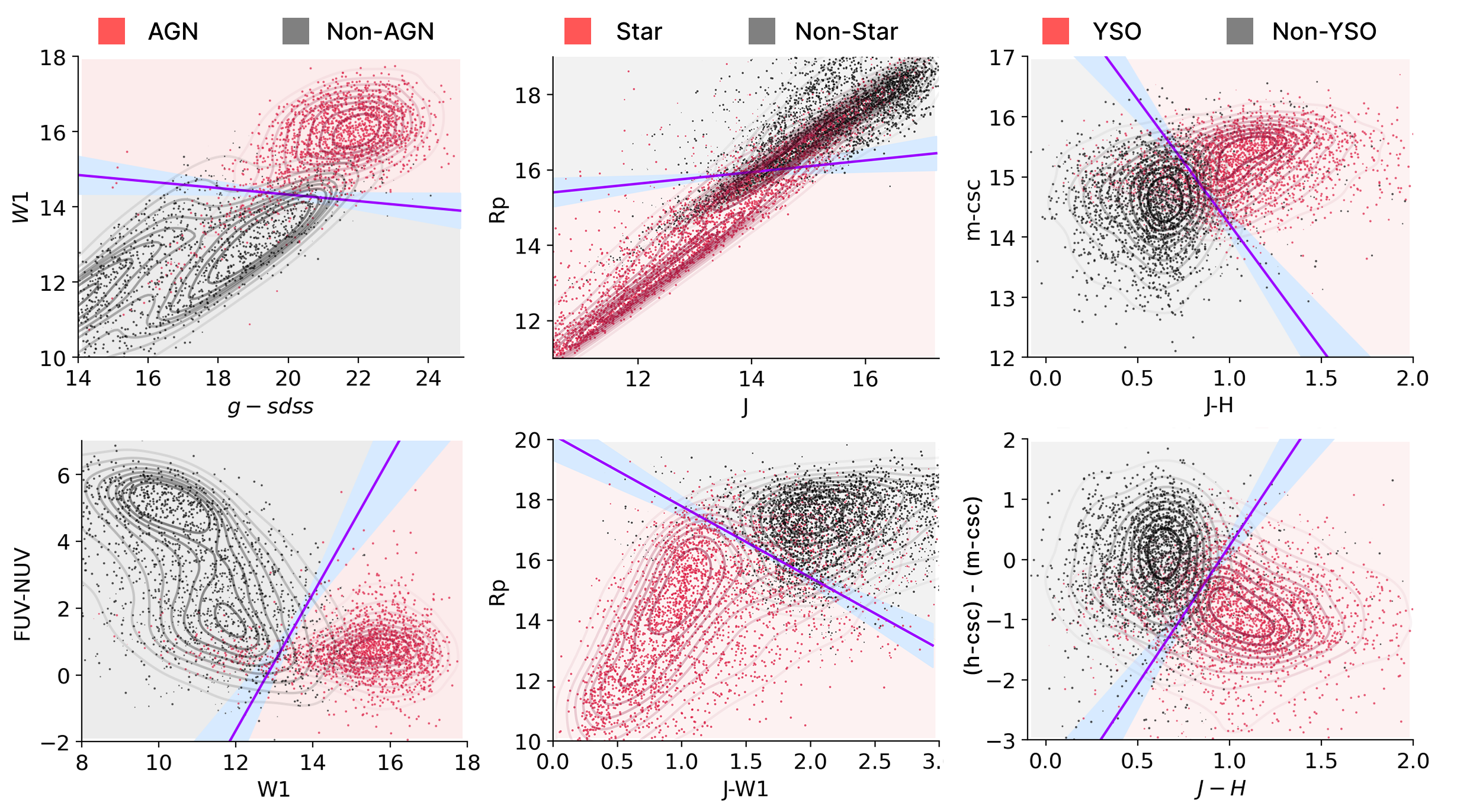}
        \caption{Scatterplot showing source clustering for positive and negative classes for the features (or their combination) picked out from the feature-feature interaction matrix. The contour lines show the probability density function, starting with 0.1 at the innermost level, and the difference between successive levels is 0.1. The decision boundary is shown with the solid magenta line and the uncertainties in a blue shade. The top panel shows clustering corresponding to the highest FFI (g-sdss vs W1 for AGN, J vs Rp for Star and J-H vs m-csc for YSO in Figure \ref{fig:feat-int}. The bottom panel shows clustering for other combinations of features picked from the interaction matrix having high FFI values. See \S\ref{sec:feature-interaction} for details.}
        \label{fig:decision-boundary}
    \end{figure*}
    
    The spread in the SHAP-feature correlation in Figure \ref{fig:shap-feat-corr} can be understood with this feature-feature interaction matrix. For example, in Figure \ref{fig:shap-feat-corr}, we see a very wide spread in the SHAP value for W1 $>$ 15. This means that even if the W1 value is identical for these sources, its importance is not the same. This must be due to the interaction of W1 with other features. In Figure \ref{fig:feat-int}, we see that the W1 has the highest interaction with the {\em SDSS} g-band, followed by FUV and NUV. Therefore, the first five global feature importance for AGNs (Figure~\ref{fig:feat-imp}) are due to their individual and feature-feature interaction contributions. For Stars, J-W1 is the top GFI (Figure~\ref{fig:feat-imp}); however, it did not show much spread in the SHAP distribution (Figure \ref{fig:shap-feat-corr}) and the same is reflected as relatively low FFI in Figure \ref{fig:feat-int}. However, the FFI is highest for the optical Rp band in Gaia and the 2MASS J band for Stars, and both these features appear in the top five GFI (Figure~\ref{fig:feat-imp}) list of Stars due to their FFI. For YSOs, the J-H interaction with m-csc and h-csc band fluxes is important, and in both cases, Figure \ref{fig:shap-feat-corr} shows a significant spread in the SHAP distribution. Also, these are among the top GFI for YSOs. This is expected for YSOs as the IR color-color and color-magnitude diagrams are generally used for YSO identification \citep{ysoIRExcess2013AJ....145..126H}. 
    
    High FFI values in the interaction matrix (\ref{fig:feat-int}) indicate that a specific pair of features exhibited a strong class-identifying relationship. To visually illustrate this, we make a scatterplot for a few selected feature pairs based on the FFI matrix (e.g. W1 and g-sdss for AGN). The top panel of Figure \ref{fig:decision-boundary} displays the scatterplot for features with the highest interaction (FFI = 1 in Figure \ref{fig:feat-int}), while the bottom panel presents a combination of other high FFI features. On the scatterplot, we label data points as per their classification outcome in the one-vs-rest classifier(refer \S \ref{sec:method}). Subsequently, we compute the class-wise two-dimensional probability density functions (PDFs) for this two-dimensional feature space. The PDFs are computed using a 2D kernel density estimate. The PDFs are visually represented in the figure by density contours, with the level of contours going from 0.1 to 1 with subsequent increments of 0.1. The class-wise PDF contours highlight the region where each class tends to cluster. Apart from the highest FFI, in the bottom panel of Figure \ref{fig:decision-boundary} we have restricted our analysis to those feature pairs where a linear decision boundary could be effectively identified. However, based on the interaction matrix, various other combinations can be explored in the online portal. To determine the linear decision boundary separating the two clusters (AGN vs. Non-AGN, Star vs. Non-Star and YSO vs. Non-YSO), we employ a Support Vector Classifier (SVC) with a linear kernel \citep{chang2011libsvm}. An SVC identifies an optimal hyperplane by maximising the distance between the hyperplane and the closest training data points (support vectors) so that the cluster boundaries are well separated. \cite{SVCexample2019MNRAS.485.1085S} used SVCs to identify linear decision boundaries in 2D and 3D feature spaces for classifying emission-line galaxies. We train an SVC using the CMP obtained from our one-versus-rest classifiers to define these decision boundaries. The SVC with a linear kernel separates the two-dimensional feature space into two distinct regions. The boundary between these two regions is taken as the decision boundary. The linear decision boundaries are shown with a magenta line in Figure \ref{fig:decision-boundary}. To compute the error (shown in light blue shade in Figure \ref{fig:decision-boundary}), we performed decision boundary calculation 500 times by randomly selecting $50\%$ of sources from both positive and negative classes each time. For a given pair of features, the decision boundary (including the associated uncertainty) can be expressed as an empirical relation, and therefore, multiple pairs of decision boundaries are useful to pick sources of a given class from a multiclass data set. Here, we discuss a few empirical relations for AGNs, Stars and YSOs based on the important FFI.

    AGN has the strongest interaction of W1 with g-sdss, and the decision boundary is:
    \begin{equation}
        W1 > -0.09\pm 0.03 \times g{\hspace{-0.03 cm}-\hspace{-0.03 cm}}sdss +16.2\pm0.5
        \label{eq:dc1}
    \end{equation}
    
    For Star, the highest interaction is between {\em Gaia} Rp and {\em 2MASS} J band.  The corresponding decision boundary is give by:
    \begin{align}
        R_p < 0.15\pm0.03 \, J + 13.7\pm0.4
    \end{align}
    
    For YSO, the J-H color has the highest and second-highest interaction with the m-csc band, with the decision boundary given as:
    \begin{align}
        m{\hspace{-0.1 cm}}-{\hspace{-0.1 cm}}csc > -4.1\pm3.4\, (J-H)+ 18.3\pm0.3
    \end{align}
    
    Although the SHAP FFI analysis is done here for the pair of features, a similar analysis can be done for any higher order of interaction. The quantification of such higher-order interactions is computationally challenging. Here we illustrate that further feature combinations also result in significant cluster separation. The bottom panel in Figure \ref{fig:decision-boundary} shows clustering and decision boundary for such additional feature combinations.
   For AGN, the {\em GALEX} FUV and NUV bands show high interaction with W1 and g-sdss bands. We find the decision boundary in the colour-magnitude diagram between FUV-NUV and W1 as : 
   \begin{align}
       FUV-NUV < 2.06\pm0.2\,W1-26.5\pm3.1
   \end{align}
   For Stars, given that Rp interaction is also higher with J-W1 color, we analyse the clustering in Rp vs J-W1 color magnitude diagram. The corresponding decision boundary is identified as :
    \begin{equation}
        R_p < -2.3\pm 0.2 (J-W1) + 20.2\pm0.3
        \label{eq:dc2}
    \end{equation}
    For YSO, \tcbr{the J-H color has the highest interaction with the m-csc band and the second-highest interaction with the h-csc band}. The decision boundary in the J-H vs h-csc - m-csc color-color diagram is given as:
    \begin{equation}
        h{\hspace{-0.05 cm}-\hspace{-0.05 cm}}csc - m{\hspace{-0.05 cm}-\hspace{-0.05 cm}}csc < 2.6\pm 0.7 (J-H) - 2.9\pm 0
        \label{eq:dc3}
    \end{equation}

    With feature interaction analysis, we understand the features and their pairs that are most effective in classifying AGNs, Stars, and YSOs, along with the decision boundary for their highest-interacting feature combinations. For AGN classification, the combination of {\em WISE} W1 band and the {\em SDSS} g-band is the most effective feature pair. The W1 magnitude, featuring as the most prominent, aligns well with the known photometric properties of AGNs, where characteristic emission in the mid-infrared due to hot dust reprocessing emission from the central engine \citep{sternAGN2012ApJ...753...30S}. Complementing this, optical surveys including the {\em SDSS} and {\em Gaia} bands are crucial for AGN selection of X-ray sources\citep{rovilos2011optical, SDSSrakshit2020spectral, gaiastorey2024quaia, xue2011chandra}.
    
    For Star classification, we identified the decision boundary in colour-magnitude diagram between {\em Gaia} R\_p band and {\em 2MASS},{\em WISE} J-W1 colour. The use of color-color diagrams in optical and IR is a well-established method for spectral classification of stars \citep{gaiacoll2018A&A...616A...1G}, and our analysis confirms the selection of this combination. For YSOs, the SHAP analysis highlights the {\em Chandra}'s medium (m) and hard (h) band combined with the {\em WISE} J-H near-infrared colour as the highest and second highest interaction features, respectively. The X-ray and optical color-color (h-csc - m-csc band with J-H) resulted in linearly separable clusters with identified decision boundary (\ref{eq:dc3}). This finding is supported by the strong emission from YSOs due to flaring activities \citep{ysoxray2007prpl.conf..313F} and the {\em 2MASS} J-H color is indicative of the IR-excess in the YSO emission \citep{ysoIRExcess2013AJ....145..126H}. 
    
    The application of SHAP analysis is crucial for identifying such relations among features and their optimal decision boundary. This level of insight is highly challenging, if not impossible, to predict by conventional means. The local explanation capability of SHAP analysis for each source allowed for interpreting the feature relations and interpretable decision boundaries, confirming that features known to be characteristic are learnt by the model. The trend figured out by the SHAP analysis with the feature's value and its importance in the outcome enhances the belief in the given source's classification. The SHAP analysis effectively unravels the relations learnt by the ML model from the feature table. The relations from this analysis are completely empirical in nature and have huge potential for giving new physical insights into the nature of the sources. \tcbr{In this work we restrict the SHAP analysis to the three majority classes (AGN, Star, YSO), which provide sufficiently large and balanced samples for robust statistics. For the minority classes (e.g., XRBs, CVs, Pulsars), the small number of confidently classified sources leads to noisy SHAP distributions that are less reliable. Nevertheless, the same methodology can in principle be applied to these classes, and our interactive catalogue framework allows users to explore SHAP values for these sources as minority classification improves in future work.}

\section{Summary and Conclusion}
    
    We present a comprehensive probabilistic classification of X-ray point sources within the \textit{Chandra} Source Catalog-2.0. Utilizing a LightGBM classifier, we successfully categorized 277,069 sources across eight astrophysical classes, including AGNs, Stars, and YSOs, with 54,770 (and 14,066 with $>4\sigma$) sources robustly classified with $>3\sigma$ confidence. For classification, multiwavelength photometric data from \textit{Chandra}, \textit{Gaia}, \textit{WISE}, 2MASS, and \textit{GALEX} are used to estimate class membership probabilities for each object.
    
To enhance the reliability,  utility and interpretability of these classifications, especially for the majority classes: AGN, Star, and YSO, we employed  SHAP analysis. We use SHAP values to derive local explanations for predictions of class membership probabilities. This allows us to explore the class-wise importance of individual MW features and their pair-wise interactions. Key findings from our SHAP analysis include:
    \begin{itemize}
        \item Most important features for identifying AGNs are the WISE W1 magnitude and the SDSS g-band. For Stars, the J–W1 and J–H color indices are most significant, while YSOs are best characterized by the J–H color and the Chandra h-csc.
    
        \item Identified Multiwavelength Thresholds: We derive thresholds for the class-wise most important feature, which has statistically contributed positively to the identification. The derived thresholds are as follows:
        \begin{itemize}
            \item AGN:  \textit{WISE} W1 magnitude $>12.5$
            \item Stars:  J-\textit{WISE} W1 color $<2.3$
            \item YSO: J-H color $>0.4$
        \end{itemize}
        \item Empirical Decision Boundaries: Analysis of feature interaction importance reveals new empirical decision boundaries that aid in distinguishing astrophysical source classes. Some of the decision boundaries are listed as follows:
        \begin{itemize}
            \item IR-optical: Between \textit{WISE} W1 and \textit{SDSS} g-band for AGN, and \textit{Gaia} Rp and 2MASS J bands for Stars.
            \item IR-X-ray: Between J-H color and \textit{Chandra} X-ray magnitude for YSOs.
        \end{itemize}
    \end{itemize}
    The SHAP explanations confirmed the model's ability to learn established identification patterns based on multiwavelength color-color and color-magnitude clustering. Crucially, this interpretability also uncovered novel empirical relations and thresholds. The astrophysical implications of these findings are substantial:
    \begin{itemize}
        \item This work provides specific, data-driven multiwavelength criteria (e.g., precise thresholds for \textit{WISE} and \textit{2MASS} magnitudes for AGN, optical/X-ray ranges for Stars, and infrared excesses for YSOs) that can guide the selection of these objects, even when complete multiwavelength data is unavailable.
        \item By making the machine learning model's reasoning transparent, we not only validate existing astrophysical selection techniques but also discover new empirical relations among MW features, which highlights the potential of using ML models to derive insights into the physical process in addition to classification.
        \item The per-source feature importance within our probabilistic classification catalogue significantly increases its value for targeted follow-up studies and area- or survey-specific investigations of newly identified \textit{Chandra} point sources.
    \end{itemize}
    While this study demonstrates the use of explainable AI for explaining the MW-based point source classification, future work will explore a more hierarchical grouping of features for all possible combinations. The probabilistic classification table alongwith local prediction explanation for individual sources will be presented as an interactive catalogue\footnote{\url{https://github.com/KumaranShivam5/Chandra-XAI.git}}. \tcbr{ Users can query the classification table and produce SHAP plots for any subset of sources or features, ensuring accessibility of the analysis} allowing the community to explore features and their combinations in greater detail. 
    
\begin{acknowledgments}

SK thanks Dr. Nilesh M. Desai (Director, SAC) and Dr. Rashmi Sharma (Deputy Director, EPSA, SAC), Dr. M.R. Pandya and Dr M.V. Shukla for their unwavering support for this work. We thank the anonyms reviewer for their comments which has greatly improved the quality and clarity of this manuscript.

 This research has made use of data obtained from the \em{Chandra} Source Catalog, provided by the \em{Chandra} X-ray Center (CXC) \dataset[DOI:10.25574/csc2]{https://doi.org/10.25574/csc2}; NASA/IPAC Extragalactic Database (NED), which is operated by the Jet Propulsion Laboratory, California Institute of Technology, under contract with the National Aeronautics and Space Administration; data from the European Space Agency (ESA) mission {\it Gaia} (\url{https://www.cosmos.esa.int/gaia}), processed by the {\it Gaia} Data Processing and Analysis Consortium (DPAC,\url{https://www.cosmos.esa.int/web/gaia/dpac/consortium}); the cross-match service provided by CDS, Strasbourg.

\end{acknowledgments}

\vspace{5mm}
\facilities{CXO}

\software{astropy \citep{astropy:2013,astropy:2018, astropy:2022},  LightGBM \citep{ke2017lightgbm, lightgbm}, scikit-learn\citep{sklearn_api}, scipy \citep{scipy}, Matplotlib \citep{Hunter:2007}}

\bibliography{referencecs}{}

@article{astropy:2013,
Adsurl = {http://adsabs.harvard.edu/abs/2013A%26A...558A..33A},
Archiveprefix = {arXiv},
Author = {{Astropy Collaboration} and {Robitaille}, T.~P. and {Tollerud}, E.~J. and {Greenfield}, P. and {Droettboom}, M. and {Bray}, E. and {Aldcroft}, T. and {Davis}, M. and {Ginsburg}, A. and {Price-Whelan}, A.~M. and {Kerzendorf}, W.~E. and {Conley}, A. and {Crighton}, N. and {Barbary}, K. and {Muna}, D. and {Ferguson}, H. and {Grollier}, F. and {Parikh}, M.~M. and {Nair}, P.~H. and {Unther}, H.~M. and {Deil}, C. and {Woillez}, J. and {Conseil}, S. and {Kramer}, R. and {Turner}, J.~E.~H. and {Singer}, L. and {Fox}, R. and {Weaver}, B.~A. and {Zabalza}, V. and {Edwards}, Z.~I. and {Azalee Bostroem}, K. and {Burke}, D.~J. and {Casey}, A.~R. and {Crawford}, S.~M. and {Dencheva}, N. and {Ely}, J. and {Jenness}, T. and {Labrie}, K. and {Lim}, P.~L. and {Pierfederici}, F. and {Pontzen}, A. and {Ptak}, A. and {Refsdal}, B. and {Servillat}, M. and {Streicher}, O.},
Doi = {10.1051/0004-6361/201322068},
Eid = {A33},
Eprint = {1307.6212},
Journal = {\aap},
Month = oct,
Pages = {A33},
Primaryclass = {astro-ph.IM},
Title = {{Astropy: A community Python package for astronomy}},
Volume = 558,
Year = 2013}

@ARTICLE{astropy:2018,
       author = {{Astropy Collaboration} and {Price-Whelan}, A.~M. and
         {Sip{\H{o}}cz}, B.~M. and {G{\"u}nther}, H.~M. and {Lim}, P.~L. and
         {Crawford}, S.~M. and {Conseil}, S. and {Shupe}, D.~L. and
         {Craig}, M.~W. and {Dencheva}, N. and {Ginsburg}, A. and {Vand
        erPlas}, J.~T. and {Bradley}, L.~D. and {P{\'e}rez-Su{\'a}rez}, D. and
         {de Val-Borro}, M. and {Aldcroft}, T.~L. and {Cruz}, K.~L. and
         {Robitaille}, T.~P. and {Tollerud}, E.~J. and {Ardelean}, C. and
         {Babej}, T. and {Bach}, Y.~P. and {Bachetti}, M. and {Bakanov}, A.~V. and
         {Bamford}, S.~P. and {Barentsen}, G. and {Barmby}, P. and
         {Baumbach}, A. and {Berry}, K.~L. and {Biscani}, F. and {Boquien}, M. and
         {Bostroem}, K.~A. and {Bouma}, L.~G. and {Brammer}, G.~B. and
         {Bray}, E.~M. and {Breytenbach}, H. and {Buddelmeijer}, H. and
         {Burke}, D.~J. and {Calderone}, G. and {Cano Rodr{\'\i}guez}, J.~L. and
         {Cara}, M. and {Cardoso}, J.~V.~M. and {Cheedella}, S. and {Copin}, Y. and
         {Corrales}, L. and {Crichton}, D. and {D'Avella}, D. and {Deil}, C. and
         {Depagne}, {\'E}. and {Dietrich}, J.~P. and {Donath}, A. and
         {Droettboom}, M. and {Earl}, N. and {Erben}, T. and {Fabbro}, S. and
         {Ferreira}, L.~A. and {Finethy}, T. and {Fox}, R.~T. and
         {Garrison}, L.~H. and {Gibbons}, S.~L.~J. and {Goldstein}, D.~A. and
         {Gommers}, R. and {Greco}, J.~P. and {Greenfield}, P. and
         {Groener}, A.~M. and {Grollier}, F. and {Hagen}, A. and {Hirst}, P. and
         {Homeier}, D. and {Horton}, A.~J. and {Hosseinzadeh}, G. and {Hu}, L. and
         {Hunkeler}, J.~S. and {Ivezi{\'c}}, {\v{Z}}. and {Jain}, A. and
         {Jenness}, T. and {Kanarek}, G. and {Kendrew}, S. and {Kern}, N.~S. and
         {Kerzendorf}, W.~E. and {Khvalko}, A. and {King}, J. and {Kirkby}, D. and
         {Kulkarni}, A.~M. and {Kumar}, A. and {Lee}, A. and {Lenz}, D. and
         {Littlefair}, S.~P. and {Ma}, Z. and {Macleod}, D.~M. and
         {Mastropietro}, M. and {McCully}, C. and {Montagnac}, S. and
         {Morris}, B.~M. and {Mueller}, M. and {Mumford}, S.~J. and {Muna}, D. and
         {Murphy}, N.~A. and {Nelson}, S. and {Nguyen}, G.~H. and
         {Ninan}, J.~P. and {N{\"o}the}, M. and {Ogaz}, S. and {Oh}, S. and
         {Parejko}, J.~K. and {Parley}, N. and {Pascual}, S. and {Patil}, R. and
         {Patil}, A.~A. and {Plunkett}, A.~L. and {Prochaska}, J.~X. and
         {Rastogi}, T. and {Reddy Janga}, V. and {Sabater}, J. and
         {Sakurikar}, P. and {Seifert}, M. and {Sherbert}, L.~E. and
         {Sherwood-Taylor}, H. and {Shih}, A.~Y. and {Sick}, J. and
         {Silbiger}, M.~T. and {Singanamalla}, S. and {Singer}, L.~P. and
         {Sladen}, P.~H. and {Sooley}, K.~A. and {Sornarajah}, S. and
         {Streicher}, O. and {Teuben}, P. and {Thomas}, S.~W. and
         {Tremblay}, G.~R. and {Turner}, J.~E.~H. and {Terr{\'o}n}, V. and
         {van Kerkwijk}, M.~H. and {de la Vega}, A. and {Watkins}, L.~L. and
         {Weaver}, B.~A. and {Whitmore}, J.~B. and {Woillez}, J. and
         {Zabalza}, V. and {Astropy Contributors}},
        title = "{The Astropy Project: Building an Open-science Project and Status of the v2.0 Core Package}",
      journal = {\aj},
         year = 2018,
        month = sep,
       volume = {156},
       number = {3},
          eid = {123},
        pages = {123},
          doi = {10.3847/1538-3881/aabc4f},
archivePrefix = {arXiv},
       eprint = {1801.02634},
 primaryClass = {astro-ph.IM},
       adsurl = {https://ui.adsabs.harvard.edu/abs/2018AJ....156..123A}
}

@ARTICLE{astropy:2022,
       author = {{Astropy Collaboration} and {Price-Whelan}, Adrian M. and {Lim}, Pey Lian and {Earl}, Nicholas and {Starkman}, Nathaniel and {Bradley}, Larry and {Shupe}, David L. and {Patil}, Aarya A. and {Corrales}, Lia and {Brasseur}, C.~E. and {N{"o}the}, Maximilian and {Donath}, Axel and {Tollerud}, Erik and {Morris}, Brett M. and {Ginsburg}, Adam and {Vaher}, Eero and {Weaver}, Benjamin A. and {Tocknell}, James and {Jamieson}, William and {van Kerkwijk}, Marten H. and {Robitaille}, Thomas P. and {Merry}, Bruce and {Bachetti}, Matteo and {G{"u}nther}, H. Moritz and {Aldcroft}, Thomas L. and {Alvarado-Montes}, Jaime A. and {Archibald}, Anne M. and {B{'o}di}, Attila and {Bapat}, Shreyas and {Barentsen}, Geert and {Baz{'a}n}, Juanjo and {Biswas}, Manish and {Boquien}, M{'e}d{'e}ric and {Burke}, D.~J. and {Cara}, Daria and {Cara}, Mihai and {Conroy}, Kyle E. and {Conseil}, Simon and {Craig}, Matthew W. and {Cross}, Robert M. and {Cruz}, Kelle L. and {D'Eugenio}, Francesco and {Dencheva}, Nadia and {Devillepoix}, Hadrien A.~R. and {Dietrich}, J{"o}rg P. and {Eigenbrot}, Arthur Davis and {Erben}, Thomas and {Ferreira}, Leonardo and {Foreman-Mackey}, Daniel and {Fox}, Ryan and {Freij}, Nabil and {Garg}, Suyog and {Geda}, Robel and {Glattly}, Lauren and {Gondhalekar}, Yash and {Gordon}, Karl D. and {Grant}, David and {Greenfield}, Perry and {Groener}, Austen M. and {Guest}, Steve and {Gurovich}, Sebastian and {Handberg}, Rasmus and {Hart}, Akeem and {Hatfield-Dodds}, Zac and {Homeier}, Derek and {Hosseinzadeh}, Griffin and {Jenness}, Tim and {Jones}, Craig K. and {Joseph}, Prajwel and {Kalmbach}, J. Bryce and {Karamehmetoglu}, Emir and {Ka{l}uszy{'n}ski}, Miko{l}aj and {Kelley}, Michael S.~P. and {Kern}, Nicholas and {Kerzendorf}, Wolfgang E. and {Koch}, Eric W. and {Kulumani}, Shankar and {Lee}, Antony and {Ly}, Chun and {Ma}, Zhiyuan and {MacBride}, Conor and {Maljaars}, Jakob M. and {Muna}, Demitri and {Murphy}, N.~A. and {Norman}, Henrik and {O'Steen}, Richard and {Oman}, Kyle A. and {Pacifici}, Camilla and {Pascual}, Sergio and {Pascual-Granado}, J. and {Patil}, Rohit R. and {Perren}, Gabriel I. and {Pickering}, Timothy E. and {Rastogi}, Tanuj and {Roulston}, Benjamin R. and {Ryan}, Daniel F. and {Rykoff}, Eli S. and {Sabater}, Jose and {Sakurikar}, Parikshit and {Salgado}, Jes{'u}s and {Sanghi}, Aniket and {Saunders}, Nicholas and {Savchenko}, Volodymyr and {Schwardt}, Ludwig and {Seifert-Eckert}, Michael and {Shih}, Albert Y. and {Jain}, Anany Shrey and {Shukla}, Gyanendra and {Sick}, Jonathan and {Simpson}, Chris and {Singanamalla}, Sudheesh and {Singer}, Leo P. and {Singhal}, Jaladh and {Sinha}, Manodeep and {Sip{H{o}}cz}, Brigitta M. and {Spitler}, Lee R. and {Stansby}, David and {Streicher}, Ole and {{{S}}umak}, Jani and {Swinbank}, John D. and {Taranu}, Dan S. and {Tewary}, Nikita and {Tremblay}, Grant R. and {Val-Borro}, Miguel de and {Van Kooten}, Samuel J. and {Vasovi{'c}}, Zlatan and {Verma}, Shresth and {de Miranda Cardoso}, Jos{'e} Vin{'i}cius and {Williams}, Peter K.~G. and {Wilson}, Tom J. and {Winkel}, Benjamin and {Wood-Vasey}, W.~M. and {Xue}, Rui and {Yoachim}, Peter and {Zhang}, Chen and {Zonca}, Andrea and {Astropy Project Contributors}},
        title = "{The Astropy Project: Sustaining and Growing a Community-oriented Open-source Project and the Latest Major Release (v5.0) of the Core Package}",
      journal = {\apj},
         year = 2022,
        month = aug,
       volume = {935},
       number = {2},
          eid = {167},
        pages = {167},
          doi = {10.3847/1538-4357/ac7c74},
archivePrefix = {arXiv},
       eprint = {2206.14220},
 primaryClass = {astro-ph.IM},
       adsurl = {https://ui.adsabs.harvard.edu/abs/2022ApJ...935..167A}
}

@article{rifkin2004defense,
  title={In defense of one-vs-all classification},
  author={Rifkin, Ryan and Klautau, Aldebaro},
  journal={Journal of machine learning research},
  volume={5},
  number={Jan},
  pages={101--141},
  year={2004}
}

@article{allwein2000reducing,
  title={Reducing multiclass to binary: A unifying approach for margin classifiers},
  author={Allwein, Erin L and Schapire, Robert E and Singer, Yoram},
  journal={Journal of machine learning research},
  volume={1},
  number={Dec},
  pages={113--141},
  year={2000}
}

@Article{Hunter:2007,
  Author    = {Hunter, J. D.},
  Title     = {Matplotlib: A 2D graphics environment},
  Journal   = {Computing in Science \& Engineering},
  Volume    = {9},
  Number    = {3},
  Pages     = {90--95},
  publisher = {IEEE COMPUTER SOC},
  doi       = {10.1109/MCSE.2007.55},
  year      = 2007
}

@Software{lightgbm,
  title = {lightgbm: Light Gradient Boosting Machine},
  author = {Yu Shi and Guolin Ke and Damien Soukhavong and James Lamb and Qi Meng and Thomas Finley and Taifeng Wang and Wei Chen and Weidong Ma and Qiwei Ye and Tie-Yan Liu and Nikita Titov and David Cortes},
  year = {2025},
  note = {R package version 4.6.0.99},
  url = {https://github.com/Microsoft/LightGBM},
}

@inproceedings{sklearn_api,
  author    = {Lars Buitinck and Gilles Louppe and Mathieu Blondel and
                Fabian Pedregosa and Andreas Mueller and Olivier Grisel and
                Vlad Niculae and Peter Prettenhofer and Alexandre Gramfort
                and Jaques Grobler and Robert Layton and Jake VanderPlas and
                Arnaud Joly and Brian Holt and Ga{\"{e}}l Varoquaux},
  title     = {{API} design for machine learning software: experiences from the scikit-learn
                project},
  booktitle = {ECML PKDD Workshop: Languages for Data Mining and Machine Learning},
  year      = {2013},
  pages = {108--122},
}

@ARTICLE{scipy,
  author  = {Virtanen, Pauli and Gommers, Ralf and Oliphant, Travis E. and
            Haberland, Matt and Reddy, Tyler and Cournapeau, David and
            Burovski, Evgeni and Peterson, Pearu and Weckesser, Warren and
            Bright, Jonathan and {van der Walt}, St{\'e}fan J. and
            Brett, Matthew and Wilson, Joshua and Millman, K. Jarrod and
            Mayorov, Nikolay and Nelson, Andrew R. J. and Jones, Eric and
            Kern, Robert and Larson, Eric and Carey, C J and
            Polat, {\.I}lhan and Feng, Yu and Moore, Eric W. and
            {VanderPlas}, Jake and Laxalde, Denis and Perktold, Josef and
            Cimrman, Robert and Henriksen, Ian and Quintero, E. A. and
            Harris, Charles R. and Archibald, Anne M. and
            Ribeiro, Ant{\^o}nio H. and Pedregosa, Fabian and
            {van Mulbregt}, Paul and {SciPy 1.0 Contributors}},
  title   = {{{SciPy} 1.0: Fundamental Algorithms for Scientific
            Computing in Python}},
  journal = {Nature Methods},
  year    = {2020},
  volume  = {17},
  pages   = {261--272},
  adsurl  = {https://rdcu.be/b08Wh},
  doi     = {10.1038/s41592-019-0686-2},
}

@ARTICLE{gaiashap2025A&A...697A.107Y,
       author = {{Ye}, Shuo and {Cui}, Wen-Yuan and {Li}, Yin-Bi and {Luo}, A. -Li and {Jones}, Hugh R.~A.},
        title = "{Deep learning interpretability analysis for carbon star identification in Gaia DR3}",
      journal = {\aap},
         year = 2025,
        month = may,
       volume = {697},
          eid = {A107},
        pages = {A107},
          doi = {10.1051/0004-6361/202449619},
archivePrefix = {arXiv},
       eprint = {2407.18754},
 primaryClass = {astro-ph.IM},
       adsurl = {https://ui.adsabs.harvard.edu/abs/2025A&A...697A.107Y}
}

@INPROCEEDINGS{ysoxray2007prpl.conf..313F,
       author = {{Feigelson}, E. and {Townsley}, L. and {G{\"u}del}, M. and {Stassun}, K.},
        title = "{X-Ray Properties of Young Stars and Stellar Clusters}",
    booktitle = {Protostars and Planets V},
         year = 2007,
       editor = {{Reipurth}, Bo and {Jewitt}, David and {Keil}, Klaus},
        month = jan,
        pages = {313},
          doi = {10.48550/arXiv.astro-ph/0602603},
archivePrefix = {arXiv},
       eprint = {astro-ph/0602603},
 primaryClass = {astro-ph},
       adsurl = {https://ui.adsabs.harvard.edu/abs/2007prpl.conf..313F}
}

@ARTICLE{gaiacoll2018A&A...616A...1G,
       author = {{Gaia Collaboration} and {Brown}, A.~G.~A. and {Vallenari}, A. and {Prusti}, T. and {de Bruijne}, J.~H.~J. and {Babusiaux}, C. and {Bailer-Jones}, C.~A.~L. and {Biermann}, M. and {Evans}, D.~W. and {Eyer}, L. and {Jansen}, F. and {Jordi}, C. and {Klioner}, S.~A. and {Lammers}, U. and {Lindegren}, L. and {Luri}, X. and {Mignard}, F. and {Panem}, C. and {Pourbaix}, D. and {Randich}, S. and {Sartoretti}, P. and {Siddiqui}, H.~I. and {Soubiran}, C. and {van Leeuwen}, F. and {Walton}, N.~A. and {Arenou}, F. and {Bastian}, U. and {Cropper}, M. and {Drimmel}, R. and {Katz}, D. and {Lattanzi}, M.~G. and {Bakker}, J. and {Cacciari}, C. and {Casta{\~n}eda}, J. and {Chaoul}, L. and {Cheek}, N. and {De Angeli}, F. and {Fabricius}, C. and {Guerra}, R. and {Holl}, B. and {Masana}, E. and {Messineo}, R. and {Mowlavi}, N. and {Nienartowicz}, K. and {Panuzzo}, P. and {Portell}, J. and {Riello}, M. and {Seabroke}, G.~M. and {Tanga}, P. and {Th{\'e}venin}, F. and {Gracia-Abril}, G. and {Comoretto}, G. and {Garcia-Reinaldos}, M. and {Teyssier}, D. and {Altmann}, M. and {Andrae}, R. and {Audard}, M. and {Bellas-Velidis}, I. and {Benson}, K. and {Berthier}, J. and {Blomme}, R. and {Burgess}, P. and {Busso}, G. and {Carry}, B. and {Cellino}, A. and {Clementini}, G. and {Clotet}, M. and {Creevey}, O. and {Davidson}, M. and {De Ridder}, J. and {Delchambre}, L. and {Dell'Oro}, A. and {Ducourant}, C. and {Fern{\'a}ndez-Hern{\'a}ndez}, J. and {Fouesneau}, M. and {Fr{\'e}mat}, Y. and {Galluccio}, L. and {Garc{\'\i}a-Torres}, M. and {Gonz{\'a}lez-N{\'u}{\~n}ez}, J. and {Gonz{\'a}lez-Vidal}, J.~J. and {Gosset}, E. and {Guy}, L.~P. and {Halbwachs}, J. -L. and {Hambly}, N.~C. and {Harrison}, D.~L. and {Hern{\'a}ndez}, J. and {Hestroffer}, D. and {Hodgkin}, S.~T. and {Hutton}, A. and {Jasniewicz}, G. and {Jean-Antoine-Piccolo}, A. and {Jordan}, S. and {Korn}, A.~J. and {Krone-Martins}, A. and {Lanzafame}, A.~C. and {Lebzelter}, T. and {L{\"o}ffler}, W. and {Manteiga}, M. and {Marrese}, P.~M. and {Mart{\'\i}n-Fleitas}, J.~M. and {Moitinho}, A. and {Mora}, A. and {Muinonen}, K. and {Osinde}, J. and {Pancino}, E. and {Pauwels}, T. and {Petit}, J. -M. and {Recio-Blanco}, A. and {Richards}, P.~J. and {Rimoldini}, L. and {Robin}, A.~C. and {Sarro}, L.~M. and {Siopis}, C. and {Smith}, M. and {Sozzetti}, A. and {S{\"u}veges}, M. and {Torra}, J. and {van Reeven}, W. and {Abbas}, U. and {Abreu Aramburu}, A. and {Accart}, S. and {Aerts}, C. and {Altavilla}, G. and {{\'A}lvarez}, M.~A. and {Alvarez}, R. and {Alves}, J. and {Anderson}, R.~I. and {Andrei}, A.~H. and {Anglada Varela}, E. and {Antiche}, E. and {Antoja}, T. and {Arcay}, B. and {Astraatmadja}, T.~L. and {Bach}, N. and {Baker}, S.~G. and {Balaguer-N{\'u}{\~n}ez}, L. and {Balm}, P. and {Barache}, C. and {Barata}, C. and {Barbato}, D. and {Barblan}, F. and {Barklem}, P.~S. and {Barrado}, D. and {Barros}, M. and {Barstow}, M.~A. and {Bartholom{\'e} Mu{\~n}oz}, S. and {Bassilana}, J. -L. and {Becciani}, U. and {Bellazzini}, M. and {Berihuete}, A. and {Bertone}, S. and {Bianchi}, L. and {Bienaym{\'e}}, O. and {Blanco-Cuaresma}, S. and {Boch}, T. and {Boeche}, C. and {Bombrun}, A. and {Borrachero}, R. and {Bossini}, D. and {Bouquillon}, S. and {Bourda}, G. and {Bragaglia}, A. and {Bramante}, L. and {Breddels}, M.~A. and {Bressan}, A. and {Brouillet}, N. and {Br{\"u}semeister}, T. and {Brugaletta}, E. and {Bucciarelli}, B. and {Burlacu}, A. and {Busonero}, D. and {Butkevich}, A.~G. and {Buzzi}, R. and {Caffau}, E. and {Cancelliere}, R. and {Cannizzaro}, G. and {Cantat-Gaudin}, T. and {Carballo}, R. and {Carlucci}, T. and {Carrasco}, J.~M. and {Casamiquela}, L. and {Castellani}, M. and {Castro-Ginard}, A. and {Charlot}, P. and {Chemin}, L. and {Chiavassa}, A. and {Cocozza}, G. and {Costigan}, G. and {Cowell}, S. and {Crifo}, F. and {Crosta}, M. and {Crowley}, C. and {Cuypers}, J. and {Dafonte}, C. and {Damerdji}, Y. and {Dapergolas}, A. and {David}, P. and {David}, M. and {de Laverny}, P. and {De Luise}, F.},
        title = "{Gaia Data Release 2. Summary of the contents and survey properties}",
      journal = {\aap},
         year = 2018,
        month = aug,
       volume = {616},
          eid = {A1},
        pages = {A1},
          doi = {10.1051/0004-6361/201833051},
archivePrefix = {arXiv},
       eprint = {1804.09365},
 primaryClass = {astro-ph.GA},
       adsurl = {https://ui.adsabs.harvard.edu/abs/2018A&A...616A...1G}
}

@article{xue2011chandra,
  title={The Chandra deep field-south survey: 4 Ms source catalogs},
  author={Xue, YQ and Luo, Bin and Brandt, WN and Bauer, FE and Lehmer, BD and Broos, PS and Schneider, DP and Alexander, DM and Brusa, Marcella and Comastri, A and others},
  journal={The Astrophysical Journal Supplement Series},
  volume={195},
  number={1},
  pages={10},
  year={2011},
  publisher={IOP Publishing}
}

@article{rovilos2011optical,
  title={Optical and infrared properties of active galactic nuclei in the Lockman Hole},
  author={Rovilos, E and Fotopoulou, S and Salvato, M and Burwitz, V and Egami, E and Hasinger, G and Szokoly, G},
  journal={Astronomy \& Astrophysics},
  volume={529},
  pages={A135},
  year={2011},
  publisher={EDP Sciences}
}

@article{SDSSrakshit2020spectral,
  title={Spectral properties of quasars from sloan digital sky survey data release 14: The catalog},
  author={Rakshit, Suvendu and Stalin, CS and Kotilainen, Jari},
  journal={The Astrophysical Journal Supplement Series},
  volume={249},
  number={1},
  pages={17},
  year={2020},
  publisher={IOP Publishing}
}

@article{gaiastorey2024quaia,
  title={Quaia, the Gaia-unWISE quasar catalog: an all-sky spectroscopic quasar sample},
  author={Storey-Fisher, Kate and Hogg, David W and Rix, Hans-Walter and Eilers, Anna-Christina and Fabbian, Giulio and Blanton, Michael R and Alonso, David},
  journal={The Astrophysical Journal},
  volume={964},
  number={1},
  pages={69},
  year={2024},
  publisher={IOP Publishing}
}

@ARTICLE{sternAGN2012ApJ...753...30S,
       author = {{Stern}, Daniel and {Assef}, Roberto J. and {Benford}, Dominic J. and {Blain}, Andrew and {Cutri}, Roc and {Dey}, Arjun and {Eisenhardt}, Peter and {Griffith}, Roger L. and {Jarrett}, T.~H. and {Lake}, Sean and {Masci}, Frank and {Petty}, Sara and {Stanford}, S.~A. and {Tsai}, Chao-Wei and {Wright}, E.~L. and {Yan}, Lin and {Harrison}, Fiona and {Madsen}, Kristin},
        title = "{Mid-infrared Selection of Active Galactic Nuclei with the Wide-Field Infrared Survey Explorer. I. Characterizing WISE-selected Active Galactic Nuclei in COSMOS}",
      journal = {\apj},
         year = 2012,
        month = jul,
       volume = {753},
       number = {1},
          eid = {30},
        pages = {30},
          doi = {10.1088/0004-637X/753/1/30},
archivePrefix = {arXiv},
       eprint = {1205.0811},
 primaryClass = {astro-ph.CO},
       adsurl = {https://ui.adsabs.harvard.edu/abs/2012ApJ...753...30S}
}

@ARTICLE{SVCexample2019MNRAS.485.1085S,
       author = {{Stampoulis}, Vasileios and {van Dyk}, David A. and {Kashyap}, Vinay L. and {Zezas}, Andreas},
        title = "{Multidimensional data-driven classification of emission-line galaxies}",
      journal = {\mnras},
         year = 2019,
        month = may,
       volume = {485},
       number = {1},
        pages = {1085-1102},
          doi = {10.1093/mnras/stz330},
archivePrefix = {arXiv},
       eprint = {1802.01233},
 primaryClass = {astro-ph.GA},
       adsurl = {https://ui.adsabs.harvard.edu/abs/2019MNRAS.485.1085S}
}

@ARTICLE{SHAPexample2025A&A...697A.107Y,
       author = {{Ye}, Shuo and {Cui}, Wen-Yuan and {Li}, Yin-Bi and {Luo}, A. -Li and {Jones}, Hugh R.~A.},
        title = "{Deep learning interpretability analysis for carbon star identification in Gaia DR3}",
      journal = {\aap},
         year = 2025,
        month = may,
       volume = {697},
          eid = {A107},
        pages = {A107},
          doi = {10.1051/0004-6361/202449619},
archivePrefix = {arXiv},
       eprint = {2407.18754},
 primaryClass = {astro-ph.IM},
       adsurl = {https://ui.adsabs.harvard.edu/abs/2025A&A...697A.107Y}
}

@article{strobl2008conditional,
  title={Conditional variable importance for random forests},
  author={Strobl, Carolin and Boulesteix, Anne-Laure and Kneib, Thomas and Augustin, Thomas and Zeileis, Achim},
  journal={BMC bioinformatics},
  volume={9},
  pages={1--11},
  year={2008},
  publisher={Springer}
}

@article{chang2011libsvm,
  title={LIBSVM: a library for support vector machines},
  author={Chang, Chih-Chung and Lin, Chih-Jen},
  journal={ACM transactions on intelligent systems and technology (TIST)},
  volume={2},
  number={3},
  pages={1--27},
  year={2011},
  publisher={Acm New York, NY, USA}
}

@misc{scot1992multivariate,
  title={Multivariate density estimation},
  author={Scot, David W},
  year={1992},
  publisher={Wiley \& Sons, New York}
}

@ARTICLE{ml-ex42021MNRAS.503.5263Z,
       author = {{Zhang}, Yanxia and {Zhao}, Yongheng and {Wu}, Xue-Bing},
        title = "{Classification of 4XMM-DR9 sources by machine learning}",
      journal = {\mnras},
         year = 2021,
        month = jun,
       volume = {503},
       number = {4},
        pages = {5263-5273},
          doi = {10.1093/mnras/stab744},
archivePrefix = {arXiv},
       eprint = {2103.08118},
 primaryClass = {astro-ph.IM},
       adsurl = {https://ui.adsabs.harvard.edu/abs/2021MNRAS.503.5263Z}
}

@ARTICLE{ml-ex32015ApJ...813...28F,
       author = {{Farrell}, Sean A. and {Murphy}, Tara and {Lo}, Kitty K.},
        title = "{Autoclassification of the Variable 3XMM Sources Using the Random Forest Machine Learning Algorithm}",
      journal = {\apj},
         year = 2015,
        month = nov,
       volume = {813},
       number = {1},
          eid = {28},
        pages = {28},
          doi = {10.1088/0004-637X/813/1/28},
archivePrefix = {arXiv},
       eprint = {1509.03714},
 primaryClass = {astro-ph.IM},
       adsurl = {https://ui.adsabs.harvard.edu/abs/2015ApJ...813...28F}
}

@ARTICLE{ml-ex-yang2022ApJ...941..104Y,
       author = {{Yang}, Hui and {Hare}, Jeremy and {Kargaltsev}, Oleg and {Volkov}, Igor and {Chen}, Steven and {Rangelov}, Blagoy},
        title = "{Classifying Unidentified X-Ray Sources in the Chandra Source Catalog Using a Multiwavelength Machine-learning Approach}",
      journal = {\apj},
         year = 2022,
        month = dec,
       volume = {941},
       number = {2},
          eid = {104},
        pages = {104},
          doi = {10.3847/1538-4357/ac952b},
archivePrefix = {arXiv},
       eprint = {2206.13656},
 primaryClass = {astro-ph.HE},
       adsurl = {https://ui.adsabs.harvard.edu/abs/2022ApJ...941..104Y}
}

@ARTICLE{ml-ex5unsupervised-csc,
       author = {{P{\'e}rez-D{\'\i}az}, V{\'\i}ctor Samuel and {Mart{\'\i}nez-Galarza}, Juan Rafael and {Caicedo}, Alexander and {D'Abrusco}, Raffaele},
        title = "{Unsupervised machine learning for the classification of astrophysical X-ray sources}",
      journal = {\mnras},
         year = 2024,
        month = mar,
       volume = {528},
       number = {3},
        pages = {4852-4871},
          doi = {10.1093/mnras/stae260},
archivePrefix = {arXiv},
       eprint = {2401.12203},
 primaryClass = {astro-ph.IM},
       adsurl = {https://ui.adsabs.harvard.edu/abs/2024MNRAS.528.4852P}
}

@ARTICLE{CSC2024ApJS..274...22E,
       author = {{Evans}, Ian N. and {Evans}, Janet D. and {Mart{\'\i}nez-Galarza}, J. Rafael and {Miller}, Joseph B. and {Primini}, Francis A. and {Azadi}, Mojegan and {Burke}, Douglas J. and {Civano}, Francesca M. and {D'Abrusco}, Raffaele and {Fabbiano}, Giuseppina and {Graessle}, Dale E. and {Grier}, John D. and {Houck}, John C. and {Lauer}, Jennifer and {McCollough}, Michael L. and {Nowak}, Michael A. and {Plummer}, David A. and {Rots}, Arnold H. and {Siemiginowska}, Aneta and {Tibbetts}, Michael S.},
        title = "{The Chandra Source Catalog Release 2 Series}",
      journal = {\apjs},
         year = 2024,
        month = oct,
       volume = {274},
       number = {2},
          eid = {22},
        pages = {22},
          doi = {10.3847/1538-4365/ad6319},
archivePrefix = {arXiv},
       eprint = {2407.10799},
 primaryClass = {astro-ph.HE},
       adsurl = {https://ui.adsabs.harvard.edu/abs/2024ApJS..274...22E}
}

@ARTICLE{2024A&A...685A.107M,
       author = {{Mechbal}, Sarah and {Ackermann}, Markus and {Kowalski}, Marek},
        title = "{Machine learning applications in studies of the physical properties of active galactic nuclei based on photometric observations}",
      journal = {\aap},
         year = 2024,
        month = may,
       volume = {685},
          eid = {A107},
        pages = {A107},
          doi = {10.1051/0004-6361/202346557},
archivePrefix = {arXiv},
       eprint = {2303.18076},
 primaryClass = {astro-ph.GA},
       adsurl = {https://ui.adsabs.harvard.edu/abs/2024A&A...685A.107M}
}

@article{lin2013classification,
  title={Classification of X-ray sources in the XMM-Newton serendipitous source catalog: Objects of special interest},
  author={Lin, Dacheng and Webb, Natalie A and Barret, Didier},
  journal={The Astrophysical Journal},
  volume={780},
  number={1},
  pages={39},
  year={2013},
  publisher={IOP Publishing}
}

@article{kim2016star,
  title={Star-galaxy classification using deep convolutional neural networks},
  author={Kim, Edward J and Brunner, Robert J},
  journal={Monthly Notices of the Royal Astronomical Society},
  pages={stw2672},
  year={2016},
  publisher={Oxford University Press}
}

@article{killestein2021transient,
  title={Transient-optimized real-bogus classification with Bayesian convolutional neural networks--sifting the GOTO candidate stream},
  author={Killestein, TL and Lyman, J and Steeghs, D and Ackley, K and Dyer, MJ and Ulaczyk, Krzysztof and Cutter, R and Mong, YL and Galloway, DK and Dhillon, V and others},
  journal={Monthly Notices of the Royal Astronomical Society},
  volume={503},
  number={4},
  pages={4838--4854},
  year={2021},
  publisher={Oxford University Press}
}

@article{xaidef10.1016/j.inffus.2019.12.012,
author = {Barredo Arrieta, Alejandro and D\'{\i}az-Rodr\'{\i}guez, Natalia and Del Ser, Javier and Bennetot, Adrien and Tabik, Siham and Barbado, Alberto and Garcia, Salvador and Gil-Lopez, Sergio and Molina, Daniel and Benjamins, Richard and Chatila, Raja and Herrera, Francisco},
title = {Explainable Artificial Intelligence (XAI): Concepts, taxonomies, opportunities and challenges toward responsible AI},
year = {2020},
issue_date = {Jun 2020},
publisher = {Elsevier Science Publishers B. V.},
address = {NLD},
volume = {58},
number = {C},
issn = {1566-2535},
url = {https://doi.org/10.1016/j.inffus.2019.12.012},
doi = {10.1016/j.inffus.2019.12.012},
journal = {Inf. Fusion},
month = {jun},
pages = {82–115},
numpages = {34}
}

@ARTICLE{xaiflare2023A&A...671A..73P,
       author = {{Panos}, Brandon and {Kleint}, Lucia and {Zbinden}, Jonas},
        title = "{Identifying preflare spectral features using explainable artificial intelligence}",
      journal = {\aap},
         year = 2023,
        month = mar,
       volume = {671},
          eid = {A73},
        pages = {A73},
          doi = {10.1051/0004-6361/202244835},
archivePrefix = {arXiv},
       eprint = {2301.01560},
 primaryClass = {astro-ph.SR},
       adsurl = {https://ui.adsabs.harvard.edu/abs/2023A&A...671A..73P}
}

@inproceedings{shap10.5555/3295222.3295230,
author = {Lundberg, Scott M. and Lee, Su-In},
title = {A unified approach to interpreting model predictions},
year = {2017},
isbn = {9781510860964},
publisher = {Curran Associates Inc.},
address = {Red Hook, NY, USA},
booktitle = {Proceedings of the 31st International Conference on Neural Information Processing Systems},
pages = {4768–4777},
numpages = {10},
location = {Long Beach, California, USA},
series = {NIPS'17}
}

@inproceedings{lime10.1145/2939672.2939778,
author = {Ribeiro, Marco Tulio and Singh, Sameer and Guestrin, Carlos},
title = {"Why Should I Trust You?": Explaining the Predictions of Any Classifier},
year = {2016},
isbn = {9781450342322},
publisher = {Association for Computing Machinery},
address = {New York, NY, USA},
url = {https://doi.org/10.1145/2939672.2939778},
doi = {10.1145/2939672.2939778},
booktitle = {Proceedings of the 22nd ACM SIGKDD International Conference on Knowledge Discovery and Data Mining},
pages = {1135–1144},
numpages = {10},
location = {San Francisco, California, USA},
series = {KDD '16}
}

@INPROCEEDINGS{grad-cam8237336,
  author={Selvaraju, Ramprasaath R. and Cogswell, Michael and Das, Abhishek and Vedantam, Ramakrishna and Parikh, Devi and Batra, Dhruv},
  booktitle={2017 IEEE International Conference on Computer Vision (ICCV)}, 
  title={Grad-CAM: Visual Explanations from Deep Networks via Gradient-Based Localization}, 
  year={2017},
  volume={},
  number={},
  pages={618-626},
  doi={10.1109/ICCV.2017.74}}

@INPROCEEDINGS{BB8237633,

  author={Fong, Ruth C. and Vedaldi, Andrea},

  booktitle={2017 IEEE International Conference on Computer Vision (ICCV)}, 

  title={Interpretable Explanations of Black Boxes by Meaningful Perturbation}, 

  year={2017},

  volume={},

  number={},

  pages={3449-3457},

  doi={10.1109/ICCV.2017.371}}

@article{XAIfleisher2022understanding,
  title={Understanding, idealization, and explainable AI},
  author={Fleisher, Will},
  journal={Episteme},
  volume={19},
  number={4},
  pages={534--560},
  year={2022},
  publisher={Cambridge University Press}
}

@ARTICLE{kauffmann2003MNRAS.346.1055K,
       author = {{Kauffmann}, Guinevere and {Heckman}, Timothy M. and {Tremonti}, Christy and {Brinchmann}, Jarle and {Charlot}, St{\'e}phane and {White}, Simon D.~M. and {Ridgway}, Susan E. and {Brinkmann}, Jon and {Fukugita}, Masataka and {Hall}, Patrick B. and {Ivezi{\'c}}, {\v{Z}}eljko and {Richards}, Gordon T. and {Schneider}, Donald P.},
        title = "{The host galaxies of active galactic nuclei}",
      journal = {\mnras},
         year = 2003,
        month = dec,
       volume = {346},
       number = {4},
        pages = {1055-1077},
          doi = {10.1111/j.1365-2966.2003.07154.x},
archivePrefix = {arXiv},
       eprint = {astro-ph/0304239},
 primaryClass = {astro-ph},
       adsurl = {https://ui.adsabs.harvard.edu/abs/2003MNRAS.346.1055K}
}

@ARTICLE{color-color2004ApJ...617..746D,
       author = {{Daddi}, E. and {Cimatti}, A. and {Renzini}, A. and {Fontana}, A. and {Mignoli}, M. and {Pozzetti}, L. and {Tozzi}, P. and {Zamorani}, G.},
        title = "{A New Photometric Technique for the Joint Selection of Star-forming and Passive Galaxies at 1.4 <\raisebox{-0.5ex}\textasciitilde z <\raisebox{-0.5ex}\textasciitilde 2.5}",
      journal = {\apj},
         year = 2004,
        month = dec,
       volume = {617},
       number = {2},
        pages = {746-764},
          doi = {10.1086/425569},
archivePrefix = {arXiv},
       eprint = {astro-ph/0409041},
 primaryClass = {astro-ph},
       adsurl = {https://ui.adsabs.harvard.edu/abs/2004ApJ...617..746D}
}

@INPROCEEDINGS{2023HEAD...2040401M,
       author = {{Martinez Galarza}, Juan},
        title = "{The Chandra Source Catalog version 2.1: New Avenues for Discovery in X-ray Datasets}",
    booktitle = {AAS/High Energy Astrophysics Division},
         year = 2023,
       series = {AAS/High Energy Astrophysics Division},
       volume = {20},
        month = sep,
          eid = {404.01},
        pages = {404.01},
       adsurl = {https://ui.adsabs.harvard.edu/abs/2023HEAD...2040401M}
}

@ARTICLE{ysoIRExcess2013AJ....145..126H,
       author = {{Huang}, Ya Fang and {Zeng Li}, Jin and {Rector}, Travis A. and {Mallamaci}, Carlos C.},
        title = "{Efficient Selection and Classification of Infrared Excess Emission Stars Based on AKARI and 2MASS Data}",
      journal = {\aj},
         year = 2013,
        month = may,
       volume = {145},
       number = {5},
          eid = {126},
        pages = {126},
          doi = {10.1088/0004-6256/145/5/126},
       adsurl = {https://ui.adsabs.harvard.edu/abs/2013AJ....145..126H}
}

@ARTICLE{agnIRexcess2011A&A...526A..86G,
       author = {{Georgantopoulos}, I. and {Rovilos}, E. and {Xilouris}, E.~M. and {Comastri}, A. and {Akylas}, A.},
        title = "{X-ray detected infrared excess AGN in the Chandra deep fields: a moderate fraction of Compton-thick sources}",
      journal = {\aap},
         year = 2011,
        month = feb,
       volume = {526},
          eid = {A86},
        pages = {A86},
          doi = {10.1051/0004-6361/201014417},
archivePrefix = {arXiv},
       eprint = {1007.0350},
 primaryClass = {astro-ph.CO},
       adsurl = {https://ui.adsabs.harvard.edu/abs/2011A&A...526A..86G}
}

@ARTICLE{chandraDeep2009ApJ...701..811T,
       author = {{Tomsick}, John A. and {Chaty}, Sylvain and {Rodriguez}, Jerome and {Walter}, Roland and {Kaaret}, Philip},
        title = "{Chandra Localizations and Spectra of Integral Sources in the Galactic Plane: The Cycle 9 Sample}",
      journal = {\apj},
         year = 2009,
        month = aug,
       volume = {701},
       number = {1},
        pages = {811-823},
          doi = {10.1088/0004-637X/701/1/811},
archivePrefix = {arXiv},
       eprint = {0906.2577},
 primaryClass = {astro-ph.HE},
       adsurl = {https://ui.adsabs.harvard.edu/abs/2009ApJ...701..811T}
}

@ARTICLE{chandraDeep2005ApJ...635..214E,
       author = {{Ebisawa}, K. and {Tsujimoto}, M. and {Paizis}, A. and {Hamaguchi}, K. and {Bamba}, A. and {Cutri}, R. and {Kaneda}, H. and {Maeda}, Y. and {Sato}, G. and {Senda}, A. and {Ueno}, M. and {Yamauchi}, S. and {Beckmann}, V. and {Courvoisier}, T.~J. -L. and {Dubath}, P. and {Nishihara}, E.},
        title = "{Chandra Deep X-Ray Observation of a Typical Galactic Plane Region and Near-Infrared Identification}",
      journal = {\apj},
         year = 2005,
        month = dec,
       volume = {635},
       number = {1},
        pages = {214-242},
          doi = {10.1086/497284},
archivePrefix = {arXiv},
       eprint = {astro-ph/0507185},
 primaryClass = {astro-ph},
       adsurl = {https://ui.adsabs.harvard.edu/abs/2005ApJ...635..214E}
}

@ARTICLE{agnWise2015ApJS..221...12S,
       author = {{Secrest}, N.~J. and {Dudik}, R.~P. and {Dorland}, B.~N. and {Zacharias}, N. and {Makarov}, V. and {Fey}, A. and {Frouard}, J. and {Finch}, C.},
        title = "{Identification of 1.4 Million Active Galactic Nuclei in the Mid-Infrared using WISE Data}",
      journal = {\apjs},
         year = 2015,
        month = nov,
       volume = {221},
       number = {1},
          eid = {12},
        pages = {12},
          doi = {10.1088/0067-0049/221/1/12},
archivePrefix = {arXiv},
       eprint = {1509.07289},
 primaryClass = {astro-ph.GA},
       adsurl = {https://ui.adsabs.harvard.edu/abs/2015ApJS..221...12S}
}

@ARTICLE{BH2023MNRAS.520.4867Q,
       author = {{Qiu}, Richard and {Ricarte}, Angelo and {Narayan}, Ramesh and {Wong}, George N. and {Chael}, Andrew and {Palumbo}, Daniel},
        title = "{Using Machine Learning to link black hole accretion flows with spatially resolved polarimetric observables}",
      journal = {\mnras},
         year = 2023,
        month = apr,
       volume = {520},
       number = {4},
        pages = {4867-4888},
          doi = {10.1093/mnras/stad466},
archivePrefix = {arXiv},
       eprint = {2212.04852},
 primaryClass = {astro-ph.HE},
       adsurl = {https://ui.adsabs.harvard.edu/abs/2023MNRAS.520.4867Q}
}

@article{lundberg2020local2global,
  title={From local explanations to global understanding with explainable AI for trees},
  author={Lundberg, Scott M. and Erion, Gabriel and Chen, Hugh and DeGrave, Alex and Prutkin, Jordan M. and Nair, Bala and Katz, Ronit and Himmelfarb, Jonathan and Bansal, Nisha and Lee, Su-In},
  journal={Nature Machine Intelligence},
  volume={2},
  number={1},
  pages={2522-5839},
  year={2020},
  publisher={Nature Publishing Group}
}

@inproceedings{NIPS2017_8a20a862,
 author = {Lundberg, Scott M and Lee, Su-In},
 booktitle = {Advances in Neural Information Processing Systems},
 editor = {I. Guyon and U. Von Luxburg and S. Bengio and H. Wallach and R. Fergus and S. Vishwanathan and R. Garnett},
 pages = {},
 publisher = {Curran Associates, Inc.},
 title = {A Unified Approach to Interpreting Model Predictions},
 url = {https://proceedings.neurips.cc/paper_files/paper/2017/file/8a20a8621978632d76c43dfd28b67767-Paper.pdf},
 volume = {30},
 year = {2017}
}

@article{lundberg2017unified,
  title={A unified approach to interpreting model predictions},
  author={Lundberg, Scott M and Lee, Su-In},
  journal={Advances in neural information processing systems},
  volume={30},
  year={2017}
}

@ARTICLE{RF2001MachL..45....5B,
       author = {{Breiman}, Leo},
        title = "{Random Forests.}",
      journal = {Machine Learning},
         year = 2001,
        month = jan,
       volume = {45},
        pages = {5-32},
          doi = {10.1023/A:1010933404324},
       adsurl = {https://ui.adsabs.harvard.edu/abs/2001MachL..45....5B}
}

@ARTICLE{2013ApJ...772...26A,
       author = {{Assef}, R.~J. and {Stern}, D. and {Kochanek}, C.~S. and {Blain}, A.~W. and {Brodwin}, M. and {Brown}, M.~J.~I. and {Donoso}, E. and {Eisenhardt}, P.~R.~M. and {Jannuzi}, B.~T. and {Jarrett}, T.~H. and {Stanford}, S.~A. and {Tsai}, C. -W. and {Wu}, J. and {Yan}, L.},
        title = "{Mid-infrared Selection of Active Galactic Nuclei with the Wide-field Infrared Survey Explorer. II. Properties of WISE-selected Active Galactic Nuclei in the NDWFS Bo{\"o}tes Field}",
      journal = {\apj},
         year = 2013,
        month = jul,
       volume = {772},
       number = {1},
          eid = {26},
        pages = {26},
          doi = {10.1088/0004-637X/772/1/26},
archivePrefix = {arXiv},
       eprint = {1209.6055},
 primaryClass = {astro-ph.CO},
       adsurl = {https://ui.adsabs.harvard.edu/abs/2013ApJ...772...26A}
}

@article{ke2017lightgbm,
  title={Lightgbm: A highly efficient gradient boosting decision tree},
  author={Ke, Guolin and Meng, Qi and Finley, Thomas and Wang, Taifeng and Chen, Wei and Ma, Weidong and Ye, Qiwei and Liu, Tie-Yan},
  journal={Advances in neural information processing systems},
  volume={30},
  pages={3146--3154},
  year={2017}
}

@ARTICLE{kumaran2023MNRAS.520.5065K,
       author = {{Kumaran}, Shivam and {Mandal}, Samir and {Bhattacharyya}, Sudip and {Mishra}, Deepak},
        title = "{Automated classification of Chandra X-ray point sources using machine learning methods}",
      journal = {\mnras},
         year = 2023,
        month = apr,
       volume = {520},
       number = {4},
        pages = {5065-5076},
          doi = {10.1093/mnras/stad414},
archivePrefix = {arXiv},
       eprint = {2302.09008},
 primaryClass = {astro-ph.HE},
       adsurl = {https://ui.adsabs.harvard.edu/abs/2023MNRAS.520.5065K}
}
\bibliographystyle{aasjournal}

\end{document}